\documentclass{jfm}

\usepackage{graphicx}
\usepackage{natbib}
\usepackage{amsmath}
\usepackage{psfrag}
\usepackage[x11names]{xcolor} % for partly colored caption

\makeatletter
\newskip\@bigflushglue \@bigflushglue = -2cm plus 1fil

\def\bigcentering{\let\\\@centercr\rightskip\@bigflushglue%
\leftskip\@bigflushglue
\parindent\z@\parfillskip\z@skip}

\ifCUPmtlplainloaded \else
  \checkfont{eurm10}
  \iffontfound
    \IfFileExists{upmath.sty}
      {\typeout{^^JFound AMS Euler Roman fonts on the system,
                   using the 'upmath' package.^^J}%
       \usepackage{upmath}}
      {\typeout{^^JFound AMS Euler Roman fonts on the system, but you
                   dont seem to have the}%
       \typeout{'upmath' package installed. JFM.cls can take advantage
                 of these fonts,^^Jif you use 'upmath' package.^^J}%
      }
  \else
  \fi
\fi

\ifCUPmtlplainloaded \else
  \checkfont{msam10}
  \iffontfound
    \IfFileExists{amssymb.sty}
      {\typeout{^^JFound AMS Symbol fonts on the system, using the
                'amssymb' package.^^J}%
       \usepackage{amssymb}%
         \let\leq=\leqslant
         
      }{}
  \fi
\fi

\ifCUPmtlplainloaded \else
  \IfFileExists{amsbsy.sty}
    {\typeout{^^JFound the 'amsbsy' package on the system, using it.^^J}%
     \usepackage{amsbsy}}
    {}
\fi

\newcommand\etal{\mbox{\textit{et al.}}}
\newcommand\ie{\mbox{\textit{i. e.}}}

\newcommand{\modif}[1]{{\color{red} #1}}
\newcommand{\removed}[1]{}
\newcommand\ic{\mathrm{i}}
\newcommand\sinc{\textrm{sinc}}

\title[Observation of resonant interactions among surface gravity waves]{Observation of resonant interactions among surface gravity waves}
\author[F. Bonnefoy \etal]%
{F. Bonnefoy$^1$\thanks{Email address for correspondence: felicien.bonnefoy@ec-nantes.fr}, F. Haudin$^2$, G. Michel$^3$, B. Semin$^3$,\break
T. Humbert$^4$, S. Auma{\^i}tre$^4$, M. Berhanu$^2$, and E. Falcon$^2$\ns
}

\affiliation{$^1$\'Ecole Centrale de Nantes, LHEEA, UMR 6598 CNRS, F-44 321 Nantes, France \\[\affilskip]
$^2$Univ. Paris Diderot, Sorbonne Paris Cit\'e, MSC, UMR 7057 CNRS, F-75 013 Paris, France\\[\affilskip]
$^3$\'Ecole Normale Sup\'erieure, LPS, UMR 8550 CNRS, F-75 005 Paris, France.\\[\affilskip]
$^4$CEA-Saclay, Sphynx, DSM, URA 2464 CNRS, F-91 191 Gif-sur-Yvette, France
}
\date{?; revised ?; accepted ?.}
\begin{document}
%\thispagestyle{empty} \setcounter{page}{}
%\date{\today}
\maketitle

\begin{abstract}
We experimentally study resonant interactions of oblique surface gravity waves in a large basin. Our results strongly extend previous experimental results performed mainly for perpendicular or collinear wave trains. We generate two oblique waves crossing at an acute angle, while we control their frequency ratio, steepnesses and directions. These mother waves mutually interact and give birth to a resonant wave whose properties (growth rate, resonant response curve and phase locking) are fully characterized. All our experimental results are found in good quantitative agreement with four-wave interaction theory with no fitting parameter. Off-resonance experiments are also reported and the relevant theoretical analysis is conducted and validated. 
\end{abstract}

\section{Introduction}
Resonant interactions between nonlinear waves are an efficient mechanism to transfer energy between scales. For instance, three-wave interactions appear in various systems involving quadratic nonlinearity such as for optical waves, hydrodynamic capillary surface waves, or elastic waves on a thin plate.

For hydrodynamic systems, experimental studies of three-wave interactions have been investigated for capillary surface waves \citep{MCG70b,HH87,haudin_16,aubourg2015}, 
% {\color{red}(Aubourg JFM 2015)} 
internal waves in stratified fluids \citep{MSW72,JMOD12} and inertial waves in fluids in rotation \citep{BMDC12}. For wave systems involving concave dispersion relation (i.e. when the wave frequency $\omega$ follows $\omega(k) \sim k^{\nu}$ with $k$ the wavenumber and $\nu<1$) or cubic nonlinearity, such as for surface gravity waves in deep-water, three-wave resonance conditions cannot be fulfilled. Four-wave interactions may then occur if interacting waves fulfill the following resonance conditions ${\mathbf k_1} + {\mathbf k_2}={\mathbf k_3} + {\mathbf k_4}$ and $\omega_1 +\omega_2= \omega_3 +\omega_4$, the angular frequencies $\omega_i$ and wave vectors ${\mathbf k_i}$ being linked by the linear wave dispersion relation $\omega_i \equiv \omega({\mathbf k_i})$. Mainly for the sake of simplicity, special attention has been given to the case of two degenerated mother waves, i.e. ${\mathbf k}_2={\mathbf k}_1$. Four-wave resonance conditions thus reduce to
\begin{equation}
\begin{cases}
2{\mathbf k_1}-{\mathbf k_3}={\mathbf k_4}\\
2\omega_1- \omega_3=\omega_4 
\end{cases} {\rm \ ,}
\label{reso}
\end{equation}
meaning that two interacting large-scale mother waves (1 and 3) can give birth to a smaller-scale daughter one (4). Hereafter, we will focus only on surface gravity waves in deep-water of linear dispersion relation
\begin{equation}
\omega({\mathbf k})=\sqrt{g|{\mathbf k}|}
\label{rdg} {\rm \ .}
\end{equation}

Four-wave interaction studies started in the early theoretical works of \citet{PhillipsJFM60} and \citet{LH62}. Surprisingly, there exists only few experiments specifically devoted to study such resonant wave interactions between water waves. \citet{LHS66} and \citet{McGoldrick66} were the first to observe the generation a daughter wave by wave interactions in the degenerated case. They notably evidenced a linear growth rate of the daughter wave, at short propagation distance, as predicted theoretically \citep{LH62}. These pioneer works were restricted to perpendicular mother waves with fixed and strong wave steepness ($ka$=0.1 with $a$ the wave amplitude) within a relatively small basin (3 m). In the same perpendicular configuration, \citet{tomita_89} confirmed the daughter growth rate to greater distances within a larger basin (54 m), still for fixed, but lower, mother-wave steepness ($ka< 0.05$). He also conducted slightly off-resonance experiments (wavenumber a few \% apart from the resonance). \modif{In all those experiments, three degenerated waves of the interacting quartet are generated mechanically (mother waves) and the fourth one (daughter wave) is growing due to four-wave interaction.} Finally, the non-degenerated case was conducted recently to observe finite amplitude effects on the resonance condition leading to persistent wave patterns \citep{hammack05,Liu2015}. \modif{In \citet{Liu2015}, an experimental investigation of steady-state resonant waves is carried out for short-crested waves. A nonlinear steady-state quartet is obtained theoretically in resonance condition by means of the homotopy analysis method. This quartet is then mechanically generated and the steady regime is indeed observed along the propagation in the basin. These experiments confirm the existence of steady-state resonant waves. In these experiments of \citet{Liu2015}, the generated wavefield consists of the four waves involved in the quartet plus some required higher order waves and therefore no daughter wave is expected in this case.} More recently, \citet{waseda15} investigated experimentally the case of resonant interactions in the presence of an underwater current. Most of these observations were supported by a dynamic model for nonlinear wave interactions \citep{Zakharov68,krasitskii94}. Note that another type of four-wave interactions involving collinear waves was extensively studied experimentally in the case of modulational instability (Benjamin-Feir instability) and focused on the growth of side-band satellites \citep{tulin-waseda-JFM99,lake77,su82,shemer99}. Such an instability is not observable in our configuration. 

Here, we performed experiments to study resonant interactions between two oblique surface gravity  waves in a large basin in the degenerated case. \modif{Like \citet{LHS66,McGoldrick66,tomita_89} we generate three mother waves of a resonant quartet and we observe the growth of the fourth wave, the daughter wave. For the first time however, our experiments are carried out  with mother waves crossing with an acute angle instead of perpendicular mother waves.} The mother-wave frequency ratio, their interaction angle and steepnesses are control parameters. We fully characterized the generation of a daughter wave for resonance conditions (growth rate, resonance response curve with angle, and phase locking between resonant waves), as well as for out-of-resonance conditions (detuning factor). All our measurements are found in quantitative agreement with four-wave interaction theory with no fitting parameter, provided that the mother-wave steepnesses are small enough ($ka< 0.1$). We also provide theoretical explanations of the phase-locking mechanism and the off-resonance detuning factor from the dynamical equations of \citet{Zakharov68}. The article is organized as follows. We first recall the resonant interaction theory, a perturbative approach only valid for short times \citep{PhillipsJFM60,LH62}, and then we present \modif{the main predictions of the dynamical equations}. Details of the derivation are given in a supplementary material. We introduce the experimental set up, report the experimental results for resonant conditions, and for out-of-resonance conditions, before drawing our conclusions.

\section{Perturbation approach of the resonant interaction theory}%\label{sec:theory}
\begin{figure*}
\begin{center}
%\vspace*{-0.5cm}
%\psfrag{a}[][]{$k_x/2k_1$}
%\psfrag{b}[][]{$k_y/2k_1$}
%\psfrag{c}[][]{$\omega_x/2\omega_1$}
%\psfrag{d}[][]{$2{\mathbf k_1}$}
%\psfrag{f}[][]{${\mathbf k_4}$}
%\psfrag{g}[][]{${\mathbf k_1}$}
%\psfrag{h}[][]{${\mathbf k_3}$}
%{\includegraphics[trim=0 30 555 390,scale=0.34]{Figures/fig01FB.eps}}
{\includegraphics[scale=0.4]{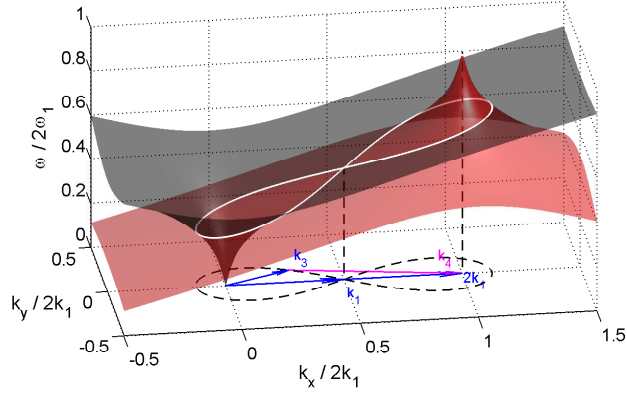}}
\caption{Solutions for four-wave resonances of surface gravity waves in the degenerated case of conditions \eqref{reso}. The dark-gray surface corresponds to $\omega({\mathbf k_3})$, \ie\ equation \eqref{rdg} with ${\mathbf k_3}=(k_x,k_y)$ and the red (light-gray) surface to the difference $2\omega({\mathbf k_1})-\omega(2{\mathbf k_1}-{\mathbf k_3})$ for a given ${\mathbf k_1}$. Resonance conditions \eqref{reso} are located on the intersection of both surfaces (white solid line). Dashed line at the bottom of the axes corresponds to the projection of the white line. Example vectors are given for $f_1=0.9$ Hz, $f_3=0.714$ Hz and $\theta=\theta_{m}=25^o$.}
\label{fig01}
\end{center}
\end{figure*}

\citet{PhillipsJFM60} and \citet{LH62} have investigated four-wave degenerated resonant solutions of \eqref{reso} for deep-water waves. A 3D representation of the solutions for a given wave vector ${\mathbf k_1}$ is shown in figure \ref{fig01} (see \citet{aubourg2015} for gravity-capillary waves). 
% The dispersion surface $\omega({\mathbf k_3})$ is defined  by equation \eqref{rdg} (black surface). Given a mother wave ${\mathbf k_1}$, a second dispersion surface is defined from degenerated resonance conditions $2\omega({\mathbf k_1})-\omega(2{\mathbf k_1}-{\mathbf k_3})$ (red surface). Intersection of both dispersion surface provides the location when the resonant conditions \eqref{reso} are fully satisfied. 
The dashed black line is exactly the classical figure-of-eight given by \citet{PhillipsJFM60}. The angle between a pair ${\mathbf k_1}$ and ${\mathbf k_3}$ on the figure of eight is noted $\theta$. The figure of eight is symmetric with respect to the ${\mathbf k_1}$ axis and either the frequency ratio $r=\omega_1/\omega_3$ or the angle $\theta$ may serve as a unique parameter to describe the eight. 
A typical example quartet is drawn in blue vectors for the mother waves and magenta for the daughter wave; it corresponds to maximal growth rate for $r=r_m=1.258$. %Note that the graphical method can be adapted easily to find resonant conditions in non-degenerated cases. 

\begin{figure*}
\begin{bigcenter}
\psfrag{a}[t][]{Angle $\theta$ ($^{\rm o}$)}
\psfrag{f}[][]{$\theta_m$}
\psfrag{g}[b][]{Growth rate $G(\theta)$}
\psfrag{v}[b][][0.5]{{$r>1$}}
\psfrag{r}[b][][0.5]{{$r<1$}}
{\includegraphics[scale=0.3]{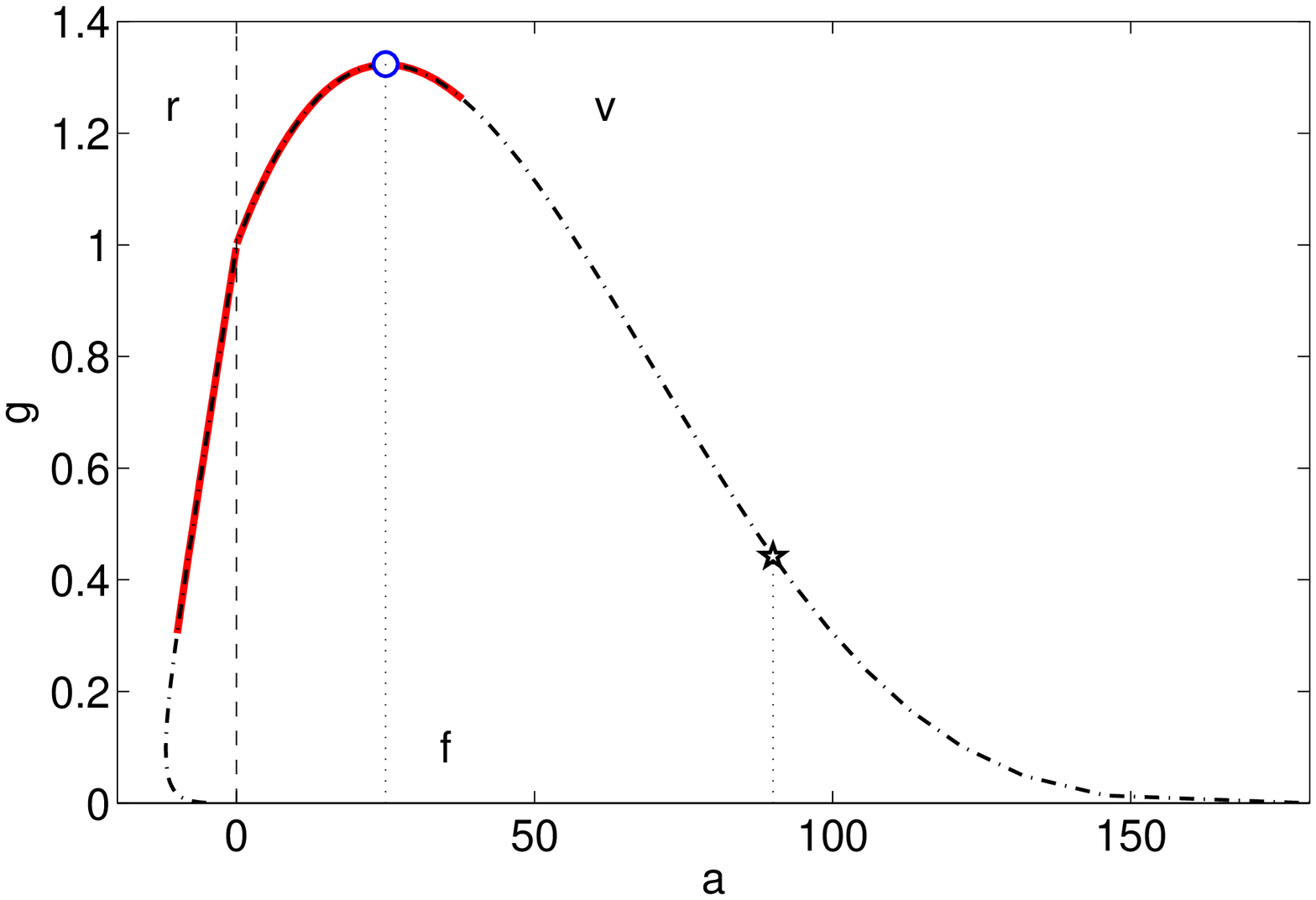}}
\psfrag{t1}[][][0.7]{{$\theta$}}
\psfrag{t2}[][][0.7]{{$\theta'$}}
\psfrag{k1}[][][0.7]{\color{green}{${\mathbf k_1}$}}
\psfrag{k3}[b][][0.7]{\color{red}{${\mathbf k_3}$}}
\psfrag{k4}[t][][0.7]{\color{blue}{${\mathbf k_4}$}}
\psfrag{k}[][][0.6]{{$k_y$}}
\psfrag{h}[][][0.6]{{$k_x$}}
\psfrag{v}[b][][0.6]{{$r>1$}}
\psfrag{r}[b][][0.6]{{$r<1$}}
{\includegraphics[scale=0.3]{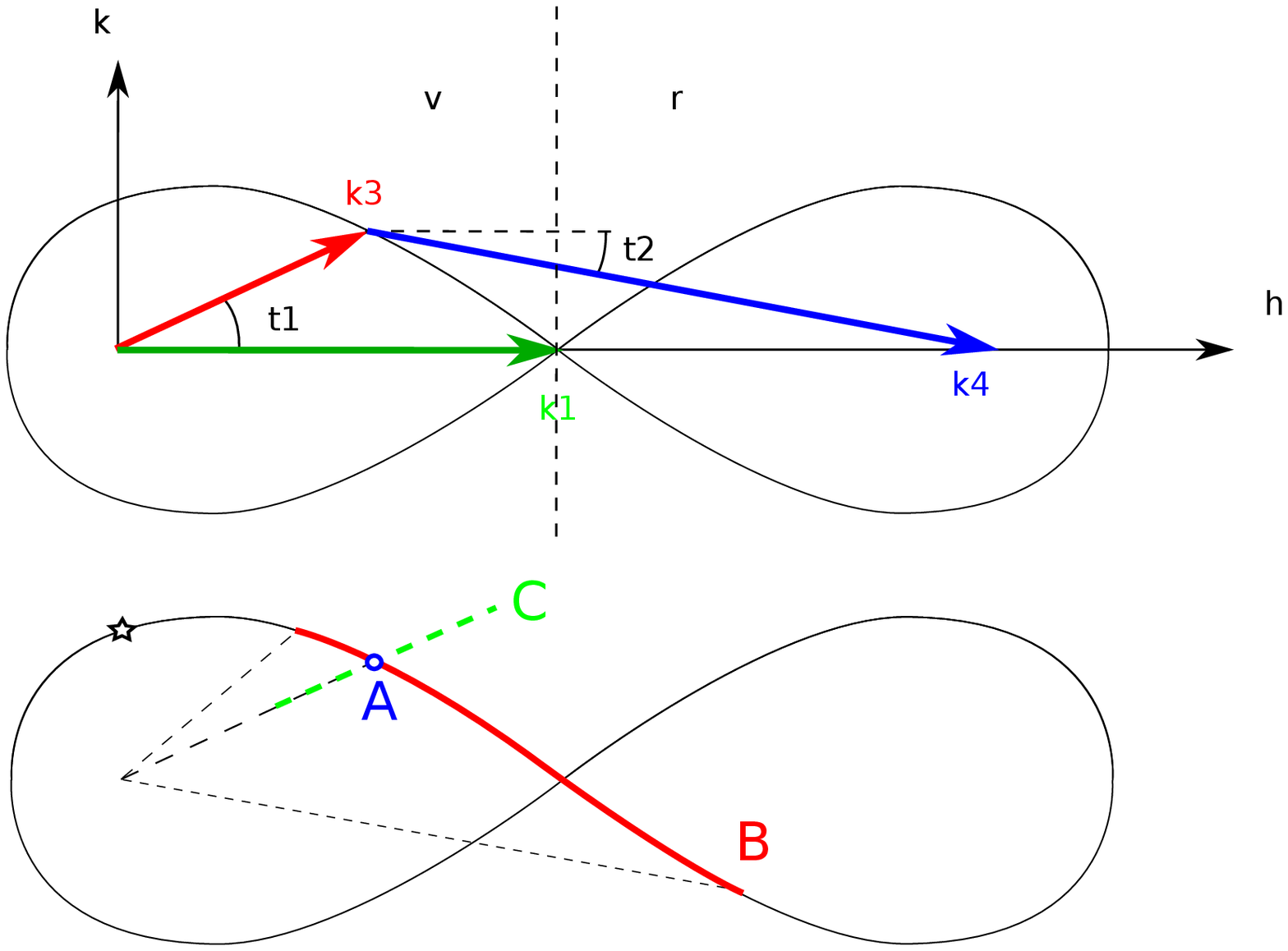}}
\caption{Left: theoretical growth-rate $G(\theta)$ of the daughter wave for degenerated case (dashed lines) and experimental tests studied in this paper: set A (blue circle), set B (red solid thick line) and experiments in litterature (black star). Top right: figure of eight with wave vectors. % and angles definition. 
Bottom right: location of the experimental tests studied in this paper: resonant experiments: same convention as in left figure with letters A and B, off-resonance experiments: set C (green dashed line).} 
\label{fig02}
\end{bigcenter}
\end{figure*}

\citet{LH62} studied theoretically the degenerated resonance in a perturbation approach considering that the mother-wave amplitudes are unaffected by the growth of the daughter wave. \citet{LH62} showed that the daughter-wave amplitude at resonance $a_4^{res}$ follows
\begin{equation}
a_4^{res} = \varepsilon_1^2 \,\varepsilon_3 \, d \, G(r) {\rm \ ,}
\label{eq:a4-LH62}
\end{equation}
where $\varepsilon_i$ are the steepnesses defined by $\varepsilon_i=k_i a_i$, $a_i$ the wave amplitude, $d$ is the distance from the wavemaker \modif{along the direction of the daughter wave} and $G$ a theoretical growth rate depending on the frequency ratio $r=\omega_1/\omega_3$. Note that the resonance conditions \eqref{reso} in deep water provide for each $r$ a unique angle $\theta$; $G$ may then be defined as a function of $r$ or $\theta$ via $r(\theta)$. The resonant daughter wave is expected to grow linearly with distance and equation \eqref{eq:a4-LH62} remains valid as long as $a_4\ll a_1$ and $a_3$. 
% We have noticed a misprint in equation (6.4) in \citet{LH62}: the term $-(6+\xi^2)^{1/2}$ should be replaced by $-{\rm sgn}(\xi)(6+\xi^2)^{1/2}$ where $\xi=(1-r)/r$. 
The growth rate $G$ is shown in figure \ref{fig02}, left, as a function of the angle $\theta$. For clarity, we have chosen positive angles for $r>1$ and negative ones for $r<1$. The growth rate is maximum for $\theta=\theta_m=25^{\rm o}$ ($r=r_m=1.258$); we locate our experimental work around this angle $\theta_m$ to obtain a significant daughter-wave amplitude; the angle $\theta$ ranges from $-10^{\rm o}$ to $+40^{\rm o}$ in our experiments. The black star on the graph of figure \ref{fig02} identifies the parameters used for the experiments of \citet{LHS66}, \citet{McGoldrick66} and \citet{tomita_89} which were all performed at $\theta=90^{\rm o}$. 

In \citet{LH62}, we can infer from the sine function describing the daughter wave and the cosine functions describing the mother waves that the phase of the daughter wave is locked to $-\pi/2$ with respect to the mother waves. 
%Such a phase-locking is also observed in other types of four-wave interaction such as the Benjamin-Feir instability \citep{benjamin_feir_67,tulin-waseda-JFM99}.

For out-of-resonance mother waves, \citet{LH62} assumes that the daughter-wave resonant growth rate is modified by a factor $\sin(\Delta kd)/\Delta kd$, which was confirmed by latter experiments \citep{LHS66,McGoldrick66}, $\Delta k$ being the wavenumber mismatch in resonance conditions \eqref{reso}. The Hamiltonian formulation given below provides a simple explanation for such a factor.

%The explanation of such a term, given in \citep{LHS66}, is based on the superposition of the near-resonant daughter wave and a free wave generated by the wavemaker with the same amplitude which is highly unlikely. The Hamiltonian formulation given below provides a simple and correct explanation.
%
\section{Hamiltonian formulation of the resonant interaction theory}\label{sec:theory}
Here, we use the framework of the approximate Hamiltonian theory of \citet{Zakharov68} with the formalism from \citet{janssen09} in order to explain the off-resonance mismatch factor. The details of the derivation are left to the supplementary material in \citet{theory}. We apply the Hamiltonian theory to a resonant degenerated interaction with two mother waves (1 and 3), present initially, and a daughter wave (4) which grows in time. \modif{The wave action amplitude is $B({\mathbf k},t)=B_1(t)\delta({\mathbf k}-{\mathbf k}_1)+B_3(t)\delta({\mathbf k}-{\mathbf k}_3)+B_4(t)\delta({\mathbf k}-{\mathbf k}_4)$ with the resonance condition $2{\mathbf k_1} - {\mathbf k_3} - {\mathbf k_4}={\mathbf 0}$ and the linear} frequency mismatch or detuning is $\Delta\omega=2\omega_1-\omega_3-\omega_4$. \modif{The Zakharov equation leads to the following evolution equation for the wave action amplitudes $B_i(t)$ of the degenerated quartet
\begin{subequations}\label{eq:zakh-degen}
\begin{align}
\ic\partial_t B_1 &= (\Omega_1-\omega_1) B_1 + 2T_{1134} \exp(\ic\Delta\omega t) B_1^* B_3 B_4 {\rm \ ,}
\label{eq:B1}\\
\ic\partial_t B_3 &= (\Omega_3-\omega_3) B_3 + \,\,\,T_{1134} \exp(-\ic\Delta\omega t) B_1^2 B_4^* {\rm \ ,}
\label{eq:B3}\\
\ic\partial_t B_4 &= (\Omega_4-\omega_4) B_4 + \,\,\,T_{1134} \exp(-\ic\Delta\omega t) B_1^2 B_3^* {\rm \ ,}
\label{eq:B4}
\end{align}
\end{subequations}
%where $\Delta\omega=2\omega_1-\omega_3-\omega_4$ is the linear frequency detuning. 
The  interaction coefficients $T_{1234} = T({\mathbf k_1},{\mathbf k_2},{\mathbf k_3},{\mathbf k_4})$ are the kernels given in \citet{krasitskii94} or \citet{janssen09}. Nonlinear frequencies $\Omega_i$ satisfy the following nonlinear dispersion relations 
\begin{equation}
\left.\begin{array}{rcl}
\Omega_1 &=& \omega_1+\,\,\,T_{1111} |B_1|^2+2T_{1313} |B_3|^2+2T_{1414} |B_4|^2 {\rm \ ,}
\\
\Omega_3 &=& \omega_3+2T_{1313} |B_1|^2+\,\,\,T_{3333} |B_3|^2+2T_{3434} |B_4|^2 {\rm \ ,}
\\
\Omega_4 &=& \omega_4+2T_{1414} |B_1|^2+2T_{3434} |B_3|^2+\,\,\,T_{4444} |B_4|^2 {\rm \ .}
\end{array}\right\}
\label{eq:NL-disp}
\end{equation}
In the early stage of the resonant interaction or for a non-resonant interaction, the daughter-wave amplitude is assumed to be negligible with respect to the mother-wave amplitudes. Equations \eqref{eq:B1} and \eqref{eq:B3} give constant magnitude and slowly evolving phase for the mother-waves while equation \eqref{eq:B4} admits the following solution
\begin{equation}
B_4 = - \ic \, T_{1134}  B_{10}^2 B_{30}^* \, \frac{\sin\left(\Delta \Omega t/2\right)}{\Delta \Omega/2}\, \exp(-\ic(\Omega_4-\omega_4+\Delta \Omega/2)t) {\rm \ .}
\label{eq:B4-sol}
\end{equation}
where the subindex 0 denotes the initial value and the total detuning is $\Delta\Omega = 2\Omega_1-\Omega_3-\Omega_4$. Derivation of this solution is straightforward and left to the supplementary material \citep{theory}. Converting to wave amplitude by means of the relation $a_i = \sqrt{2k_i/\omega_i} \, B_i$, we can infer the following wave solutions.}

At short time when $|a_4|\ll |a_{10}|,\,|a_{30}|$, we obtain constant mother amplitudes $a_i(t)=a_{i0}$ (subindex 0 means initial value). The daughter-wave amplitude \modif{and phase are} 
\begin{subequations}\label{eq:a4-mag-phase}
\begin{align}
|a_4| &= T_{1134} \frac{\omega_1}{2k_1^3} \sqrt{\frac{\omega_3 \, k_4}{\omega_4 \, k_3^3}} \varepsilon_{1}^2 \varepsilon_{3} \, \left|\frac{\sin\left(\Delta \Omega t/2\right)}{\Delta \Omega/2}\right|{\rm \ ,}
\label{eq:a4-mag}\\
\arg a_4 &= -\frac{\pi}{2} + 2\arg a_{10}-\arg a_{30} - (\Omega_4-\omega_4+\Delta \Omega/2)t{\rm \ ,}
\label{eq:a4-phase}
\end{align}
\end{subequations}
where the steepness is defined by its initial value $\varepsilon_{i}=k_i|a_{i0}|$\removed{ and the total detuning is $\Delta\Omega = 2\Omega_1-\Omega_3-\Omega_4$, $\Omega_i$ being the nonlinear frequencies given by the nonlinear dispersion relations (see \citep{theory} for details). The interaction coefficients $T_{1134} = T({\mathbf k_1},{\mathbf k_1},{\mathbf k_3},{\mathbf k_4})$ may be found in \citet{krasitskii94} or \citet{janssen09} }. Equation \eqref{eq:a4-mag} provides the evolution of the daughter-wave amplitude while equation \eqref{eq:a4-phase} gives the nonlinear evolution of its phase. % Detail analysis of this solution is given below, for resonant waves and off-resonance waves.

At resonance ($\Delta \omega=0$) and at short time ($\Delta \Omega t\ll 1$), we have $\sin\left(\Delta \Omega t/2\right) / (\Delta \Omega/2)\simeq t$. Equation \eqref{eq:a4-mag} now becomes $|a_4^{res}|=T_{1134} \omega_1 \sqrt{\omega_3 \, k_4} / (2k_1^3\sqrt{\omega_4 \, k_3^3}) \varepsilon_{10}^2 \varepsilon_{30} t$ which corresponds to the same results as in \citet{LH62}. Equation \eqref{eq:a4-phase} shows that the daughter wave phase is phase-locked to $\arg a_{40}=-\pi/2+2\arg a_{10}-\arg a_{30}$.%,  where the subindex 0 denotes the initial value.

In the case of mechanically generated mother waves, 
%the wave field consists of the two generated mother waves 1 and 3 plus the daughter wave. The later 
the daughter-wave frequency follows from exact resonance condition $\omega_4=2\omega_1-\omega_3$. It is necessary to replace time $t$ in equations \eqref{eq:a4-mag-phase} by $d/c_{g4}$ where $c_{g4}$ is the group velocity of the daughter wave and $d$ the distance in the daughter-wave direction. All the following results are valid in the steady regime between the wavemaker and the daughter-wave front. At resonance, the theoretical amplitude of the resonant wave along the basin is the same as in equation \eqref{eq:a4-LH62} (the link between $G$ and $T_{1134}$ is given in the supplementary material). 

We consider now an off-resonance degenerated quartet with a linear frequency detuning $\Delta_{}\omega\neq 0$. At the early stage of the interaction when the daughter amplitude is small compared to the mother ones, expression \eqref{eq:a4-mag} shows that the daughter amplitude evolves as a sine function. We may rewrite equation \eqref{eq:a4-mag} as $|a_4|=|a_4^{res}| \sinc \Delta \Omega t/2$. Note that this mismatch factor involves the total detuning $\Delta \Omega$ which consists of both linear and nonlinear components. At longer time, the phase mismatch will change from its initial $\Delta \omega$ value due to nonlinear dispersion. For off-resonant mechanically generated mother waves, the direction $\theta_4$ of the daughter wavenumber ${\mathbf k_4}$ is yet unknown; the condition for wavenumbers is not fullfilled and a wavevector mismatch exists, $\Delta {\mathbf k}=2{\mathbf k_1}-{\mathbf k_3}-{\mathbf k_4}$. Although the direction of the daughter wave is not specified, we assume that the fastest growing daughter wave is the one with minimal detuning. In other words, the daughter wave propagates along the direction of $2{\mathbf k_1}-{\mathbf k_3}$ and the corresponding mismatch is now $\Delta k=\left|2{\mathbf k_1}-{\mathbf k_3}\right|-k(2\omega_1-\omega_3)$. From equation \eqref{eq:a4-mag}, the off-resonance amplitude of the daughter wave is given by the same expression as in \citet{LH62}
\begin{equation}
a_4 = \varepsilon_1^2\varepsilon_3 d \, G(r,\theta_{}) \left|\frac{\sin{\frac{1}{2} \Delta k d}}{\frac{1}{2} \Delta k d}\right| = a_4^{res} \, \left|\textrm{sinc} \, \frac{\Delta k d}{2} \right| {\rm \ .}
\label{eq:eps4noreso}
\end{equation}
%This expression is the same as in \citet{LH62}. 
Note that the nonlinear detuning terms have been omitted here for clarity.
\section{Experimental setup}
The experiments presented here are designed to test the resonance theory for wave directions different from the perpendicular case studied in the 60s and by \citet{tomita_89}. We mechanically generate bichromatic waves (mother waves 1 and 3) in a rectangular wave basin and observe the birth of the daughter wave of frequency $2\omega_1-\omega_3$ due to resonant interaction (see the supplementary movie available online at doi: 10.1017/jfm..).
\begin{figure*}
\begin{bigcenter}
\psfrag{x}[l][][1]{{$\theta_{4m}$}}
\psfrag{th}[][][1]{{$\theta$}}
\psfrag{k1}[][][1]{\color{green}{${\mathbf k_1}$}}
\psfrag{k3}[l][][1]{\color{red}{${\mathbf k_3}$}}
\psfrag{k4}[][][1]{\color{blue}{${\mathbf k_4}$}}
{\includegraphics[scale=0.42]{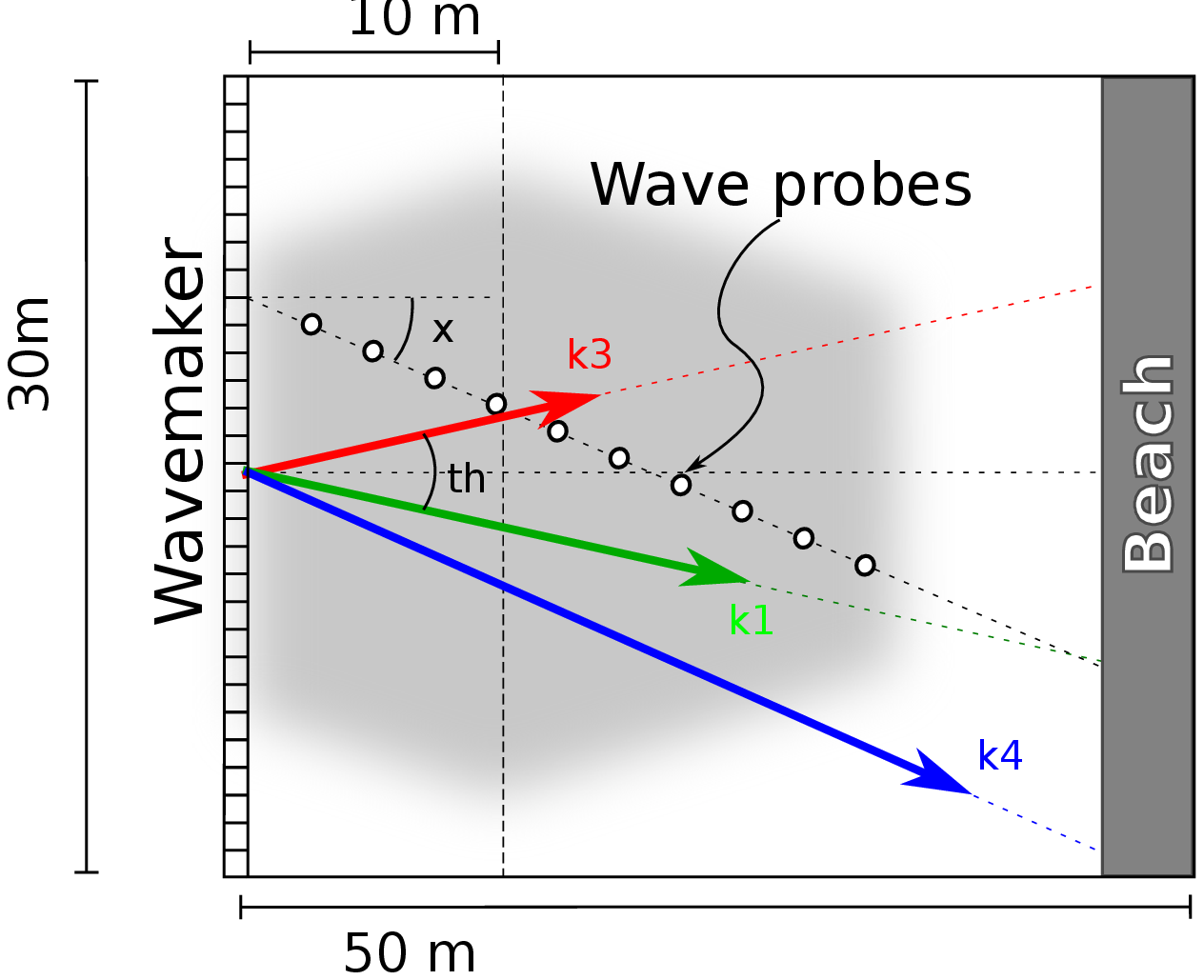}}
\hspace{0.8cm}
%\psfrag{a}[][][0.6]{{$f_1$}}
%\psfrag{b}[][][0.6]{{$f_3$}}
%\psfrag{c}[][][0.6]{{$f_4=$\newline $2f_1-f_3$}}
%\psfrag{d}[][][0.6]{{$2f_3$}}
%\psfrag{e}[][][0.6]{{$2f_1$}}
%\psfrag{f}[][][0.6]{{$f_1+f_3$}}
{\includegraphics[scale=0.4]{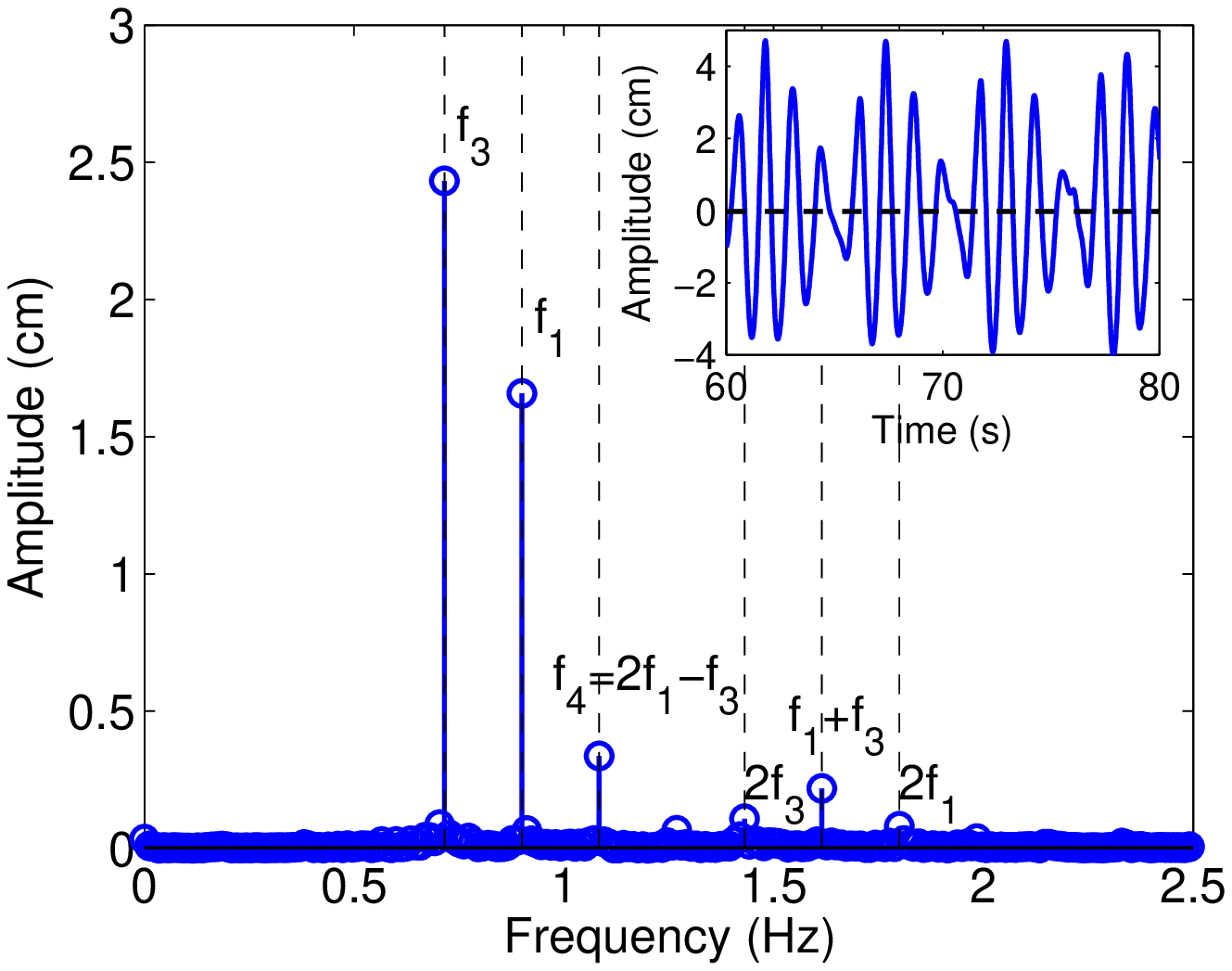}}
\caption{Left: Wave basin showing the homogeneous zone (shaded area), the wave probes (circles) and the wave vectors ${\mathbf k_1}$, ${\mathbf k_3}$ and ${\mathbf k_4}$ for the maximum growth rate case (arrows resp. in green, red and blue), right: Frequency spectrum of wave height $a(t)$ recorded at $d=21.5$ m. Vertical dashed lines correspond to frequencies: $f_3$, $f_1$, $f_4$, $2f_3$, $f_1+f_3$, and $2f_1$. Inset: Temporal evolution of the wave height, $a(t)$, dashed line is $\langle a \rangle_t \simeq 0$. Wave conditions $r=r_m$, $\theta_{}=\theta_{m}$ and $\varepsilon_1=\varepsilon_3=0.05$} 
% info: sonde 10, C7_test_0606
\label{fig03}
\end{bigcenter}
\end{figure*}
%\caption{Frequency spectrum of the recorded wave height $a(t)$. Vertical dashed lines correspond to frequencies: $f_3$, $f_1$, $f_4$, $2f_3$, $f_1+f_3$, and $2f_1$. Inset: Temporal evolution of the wave height, $a(t)$, recorded at $x=21.5$ m . Dashed line is $\langle a \rangle_t \simeq 0$. Wave conditions $r=r_m$, $\theta_{}=\theta_{m}$ and $\varepsilon_1=\varepsilon_3=0.05$.} 
The wave basin at Ecole Centrale de Nantes has dimensions 50 m $\times$ 30 m $\times$ 5 m and its wavemaker consists of independant 48 flaps that are hinged 2.8 m below the free surface. Figure \ref{fig03}, left, shows a top view of the setup. In order to avoid spurious reflections on the side-walls, the motion of the  segmented wavemaker is controlled by means of the Dalrymple method \citep{dalrymple-method-89}. The Dalrymple method aims at generating the target wave field at a distance $X_d=10$ m from the wavemaker and yields a quasi-uniform wave field from the wavemaker up to 25 m (see the grey zone of figure \ref{fig03}); this is crucial for these interaction experiments. 

The input parameters to the wavemaker are mother-wave frequency ($f_1$ and $f_3$), steepness (or amplitude $a_1$ or $a_3$) and direction ($\theta_1$ and $\theta_3$ with respect to the basin main axis).  The daughter wave direction is defined as %$\theta'$ (see figure \ref{fig02}, left) with respect to the mother wave 1 or as 
$\theta_4$ in the wave basin% ($\theta_4=\theta_1+\theta'$)
.   %Mother-waves phase is set to zero at origin which is in space the right-corner of the basin when looking downstream from the wavemaker and in time the begin of the wavemaker motion. 
Frequencies for the mother waves are chosen to fit the basin capacities: fixed $f_1=0.9$ Hz (wavelength $\lambda_1\simeq2$ m) and varied $f_3=f_1/r$ with $r=0.8$ to 1.6. The corresponding wavelengths $\lambda_3$ ranged from 1.3 to 4 m. The angle $\theta=\theta_3-\theta_1$ between mother waves 1 and 3 was varied between -10 and 40$^{\rm o}$ with a focus at $\theta_{m}=$25$^{\rm o}$ where the maximum growth rate of the daughter wave occurs ($r_m=1.258$, see figure \ref{fig02}). In this case, we have %$\theta'_m=-10.6^{\rm o}$ and the direction of the daughter wave is $\theta_{4m}=\theta_{1m}+\theta'_m$.
$\theta_{4}=\theta_{4m}=-23.1^{\rm o}$.

Three sets of experiments are presented in the following, two at resonance and one out-of-resonance.  In the first set of experiments, (set A correspond to the point A in figure \ref{fig02} right), the scaling of the daughter-wave steepness $\varepsilon_4$ is tested by varying $\varepsilon_1\in [0.01;0.1]$ at the resonance condition with maximum growth rate (that is $r=r_m$) and for fixed $\varepsilon_3=0.05$. In set B, the figure-of-eight is tested in the range $\theta_{}\in [-10^{\rm o};40^{\rm o}]$, for fixed steepnesses $\varepsilon_1=\varepsilon_3=0.07$. This corresponds to the red line on the figure of eight in figure \ref{fig02}, right. Finally, in set C, we study out-of-resonance conditions by fixing $f_1=0.9$ Hz and $\theta=\theta_m$ but changing $k_3$ by varying $r\in [1.1;1.6]$ around $r_m$, again with fixed steepnesses $\varepsilon_1=\varepsilon_3=0.05$. This corresponds to the dashed green line in figure \ref{fig02}, right. 
%Note that references to mother-wave steepnesses are the initial values; in other words, $\varepsilon_i$ for $i=1$, 3 means $\varepsilon_{i0}$.

For cases A and C, wave directions in the basin are made symmetrical $\theta_1=-\theta_{m}/2$ and $\theta_3=\theta_{m}/2$ to maximize the uniformity of the wave field. \modif{The direction of the daughter wave is $\theta_{4m}=-23.1^{\rm o}$ which corresponds to theses cases A and C with maximum growth rate when $\theta=\theta_{m}$. A linear frame supporting an array of twelve resistive wave probes is setup in the direction $\theta_{4m}$ (see figure \ref{fig03}, left). The distance between two successive probes is about 2 m. In all experiments, this linear array of wave probes is indeed aligned along the direction of the daughter wave $\theta_{4m}=-23.1^{\rm o}$. The distance $d$ to the wavemaker and measured along the direction of the daughter wave is ranging from $d=$2.5 to 25 m.}

For case B, the directions of the mother waves $\theta_1$ and $\theta_3$ were chosen in such a way that the target angle $\theta$ is obtained and that the daughter wave is aligned with the probe array.

The sampling frequency is 100 Hz. Wave heights were recorded during about 100 s which corresponds to steady regime of more than 50 wave periods. Typical amplitudes are $a_{1,3}\simeq$ few cm for mother waves and $a_4\simeq$ few mm for daughter waves. 
\section{Resonant wave conditions}
\begin{figure*}
\begin{bigcenter}
\psfrag{d}[t][]{Distance $d$ (m)}
\psfrag{a}[b][]{Amplitude $a_4$ (cm)}
{\includegraphics[scale=0.42]{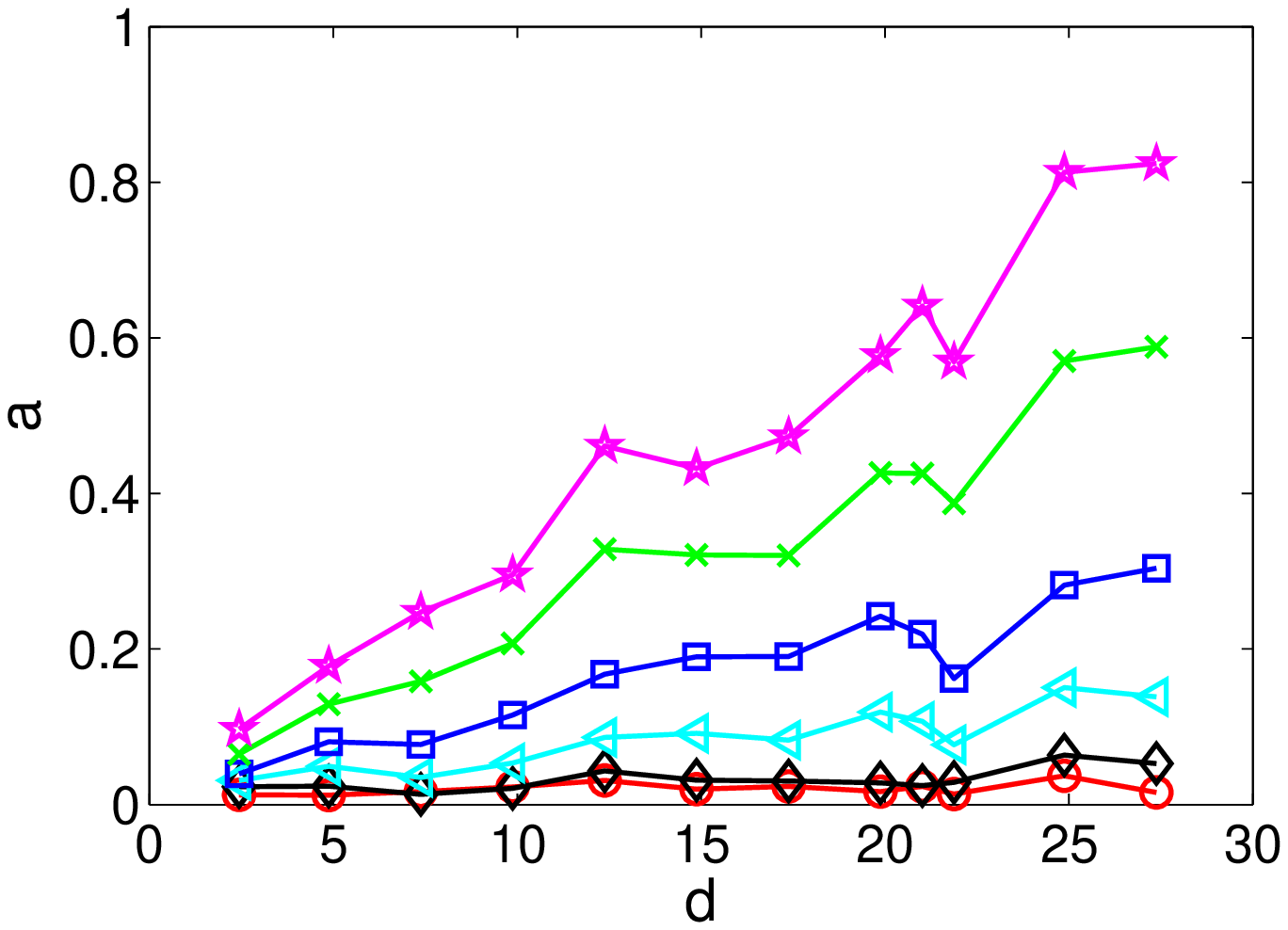}}
\psfrag{e}[t][]{$\varepsilon_1^2$}
\psfrag{b}[b][]{$a_4/[d\varepsilon_3 G(\theta_m)]$}
{\includegraphics[scale=0.42]{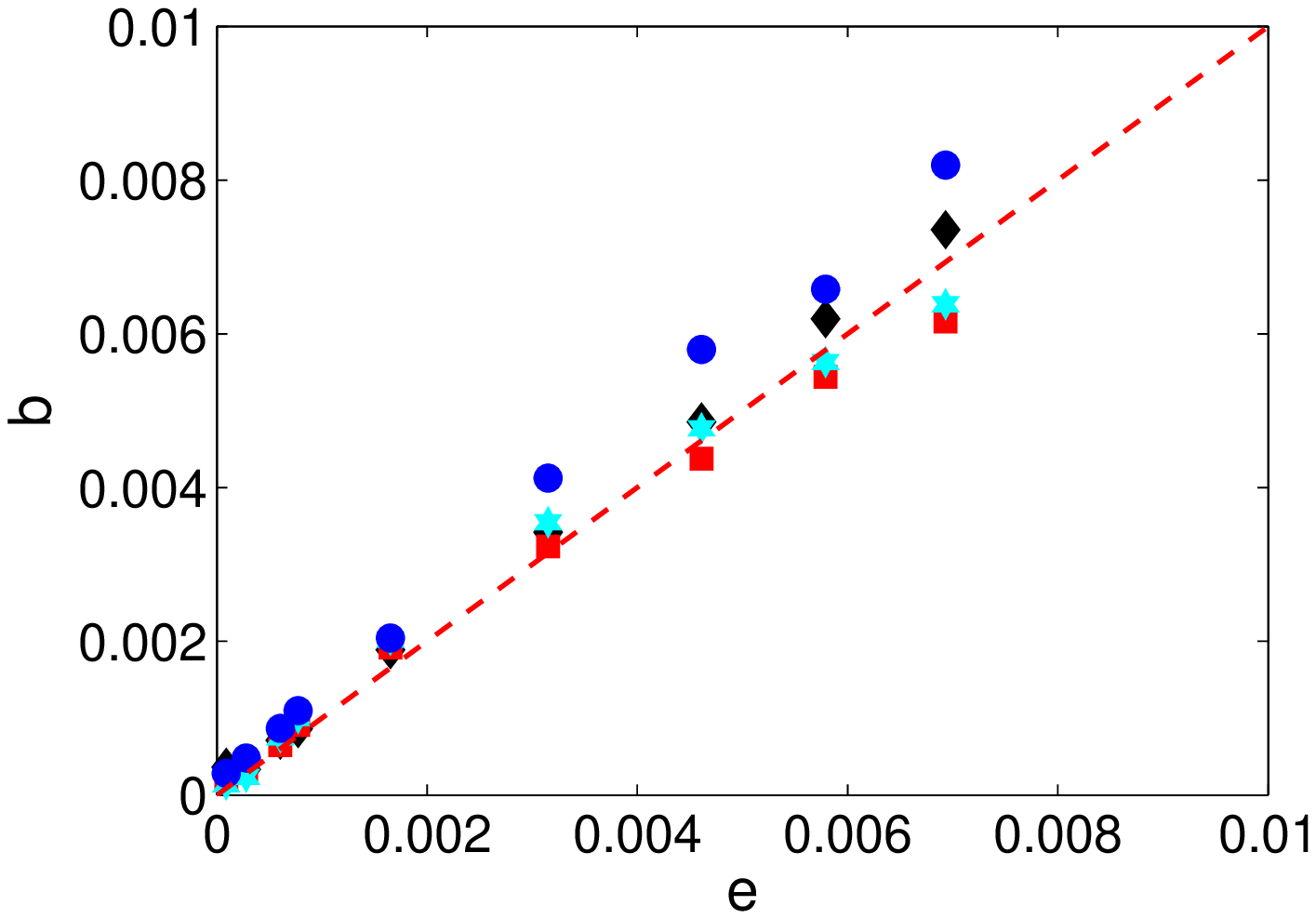}}
\caption{Amplitude of the resonant wave $a_4$ for $\varepsilon_3=0.05$ and $r=r_m$. Left: amplitude $a_4$ versus distance, $d$, for different $\varepsilon_1\times10^{3}=10$, 17, 28, 41, 56, 68 (from bottom to top). Right: rescaled amplitude of the resonant wave $a_4/[d\varepsilon_3 G(\theta_m)]$ as a function of $\varepsilon_1^2$ for different distances $d=9.9$ ($\blacklozenge$), 14.9 ($\blacksquare$), 19.9 ($\ast$), and 24.9 ($\bullet$) m. The dashed line of unity slope is  expected from equation \eqref{eq:a4-LH62}. }
% \todo{Figure de gauche \`a refaire en virant la derni\`ere sonde qui est en-dehors de la zone homog\`ene. }
% info  - serie12C6
\label{fig04}
\end{bigcenter}
\end{figure*}
% \caption{Rescaled amplitude of the resonant wave $a_4$ as a function of $\varepsilon_1^2$ for different distances $d=9.9$ ($\blacklozenge$), 14.9 ($\blacksquare$), 19.9 ($\ast$), and 24.9 ($\bullet$) m. Dashed lines has a unity slope. $\varepsilon_3=0.05$. $r=r_m$ - serie12C6. Inset: unrescaled amplitude $a_4$ versus distance, $d$, for different $\varepsilon_1=0.01$ up to 0.07 (from bottom to top).}

We report here our results for resonant degenerated quartets near maximum amplification (case A). A typical example of a temporal evolution of wave elevation $a(t)$ recorded by a probe is shown in the inset of figure \ref{fig03}, right. From the time-series measured at the wave probes, we select a steady-state window after the wave front passed the probe (time window is more than 50 periods long). A Discrete Fourier Transform is applied to the windowed signal with a standard FFT algorithm (frequency resolution is below 20 mHz). The main figure \ref{fig03}, right, shows the corresponding amplitude spectrum for case A. The two mother waves were visible at frequency $f_1$ and $f_3$. The peak at frequency $f_4=2f_1-f_3$ confirms the existence of the daugther wave, but, as expected, its amplitude is smaller than the mother-wave ones. This is a first evidence of a daughter wave generated by resonant interaction. 
% After a distance of 21.5 m, the measured daughter-wave amplitude (3.4 mm) was in very good agreement with the theoretical amplitude (3.5 mm). 
Note that harmonics at frequency $2f_3$, $f_1+f_3$ and $2f_1$ are also visible, with amplitudes yet lower than that of the daughter wave. They are the signature of second order bound waves accompanying the mother waves. The harmonics at $3f_3$ and $2f_3-f_1$ corresponding to the third order bound waves are barely visible. 

Figure \ref{fig04} left shows the daughter-wave amplitude $a_4$ as a function of distance $d$ for different steepnesses. This amplitude is found to grow linearly with distance $d$ as expected from equation \eqref{eq:a4-LH62} and to increase with the mother-wave steepness $\varepsilon_1$. Note that the experiments when $\varepsilon_1$ is fixed and $\varepsilon_3$ is varied (not shown here) show that the daughter amplitude $a_4$ grow linearly with $\varepsilon_3$ as predicted. The rescaled daughter-wave amplitude $a_4/(\varepsilon_3dG(\theta_m))$ is then shown in figure \ref{fig04} (right) as a function of $\varepsilon_1^2$ at different distances $d$. A good quantitative agreement with the theoretical predictions of equation \eqref{eq:a4-LH62} is observed, with no fitting parameter. 

\begin{figure*}
\begin{bigcenter}
\psfrag{t}[t][]{Time $t$ (s)}
\psfrag{b}[b][]{$\sin \varphi_i$, $i=1$ to 4}
{\includegraphics[scale=0.4]{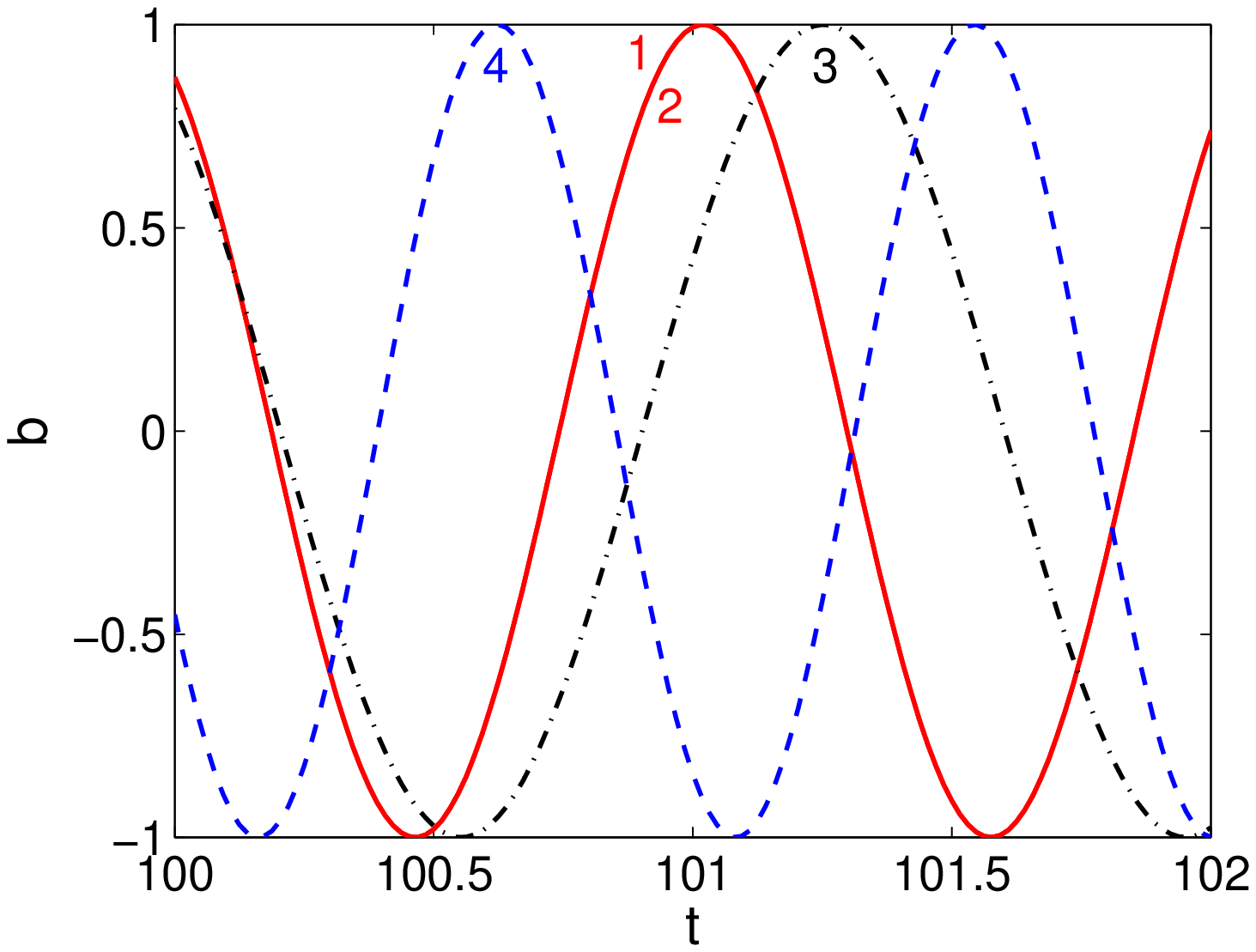}}
\psfrag{a}[b][]{$\sin (2\varphi_1-\varphi_3-\varphi_4)$}
{\includegraphics[scale=0.4]{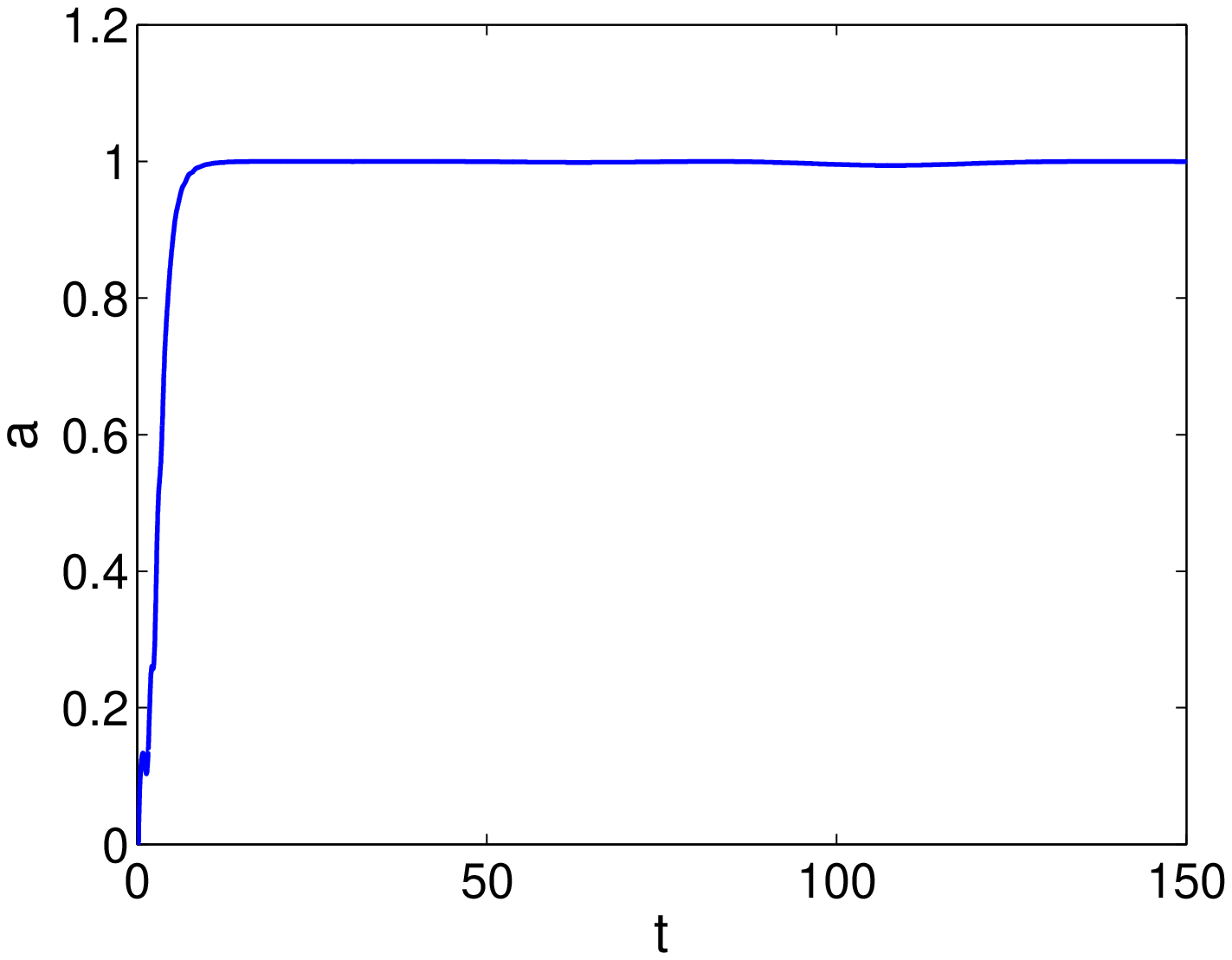}}
\caption{Left: Temporal evolution of individual phase $\varphi_i(t) \equiv {\mathbf k_i}.{\mathbf x_p}-\omega_i t+\varphi_{i0}$ of each wave $i=1$ ($-$), 3 ($.-$), and 4 ($--$). Right: Temporal evolution of the sine of the interaction phase $\varphi(t)=2\varphi_1-\varphi_3-\varphi_4$. At resonance, the latter reduces to $2\varphi_{10} -\varphi_{30}-\varphi_{40}$ which is constant (phase-locking) equal to $\pi/2$ during the experiment. Conditions $r=r_m$, $\varepsilon_1=\varepsilon_3=0.05$ at distance $d=$21.5 m.}
% info C7test0606
\label{fig05}
\end{bigcenter}
\end{figure*}
% \caption{Temporal evolution of the sum of wave phases $2\Phi_1-\Phi_3-\Phi_4$. The latter reduces to $2\varphi_1 -\varphi_3-\varphi_4=-\pi/2$ when phase sum are locked. Subinset: Temporal evolution of phase $\Phi_i \equiv k_i.x-\omega_i t+\varphi_i$ of each wave $i=1$ ($-$), 3 ($.-$), and 4 ($--$). $r=r_m$. $\varepsilon_1=\varepsilon_3=0.05$. $d=$sonde\#10. C7test0606}
%

For a given probe at the far end of the homogeneous zone, we separate the two mother waves and the daughter wave with appropriate bandpass filters around each component $f_1$, $f_3$ and $2f_1-f_3$. To wit, we compute the Hilbert transform of each component and we obtain the wave envelope $a_i(t)$ and instantaneous wave phase $\varphi_i(t) \equiv {\mathbf k_i}.{\mathbf x_p}-\omega_i t+\varphi_{i0}$, where ${\mathbf x_p}$ is the probe position. The phase of each wave $\varphi_i(t)$ is shown in the left of figure \ref{fig05} and obviously changes with time. On the contrary, the interaction phase defined by $\varphi(t)=2\varphi_1(t)-\varphi_3(t)-\varphi_4(t)$ is constant with time, as shown in figure \ref{fig05}, right. After the wave front has passed the probes, the interaction phase $\varphi$ is locked at $\pi/2$. This phase-locking demonstrated by our experiments is in very good agreement with the phase-locking predicted by equation \eqref{eq:a4-phase} for short distance (\ie\ $a_4\ll a_1$ and $a_3$). 
%The value of $\pi/2$  means that the energy transfer is maximum from wave 1 to wave 4 as shown by Equation \eqref{eq:amp4}. %This is the value of the daughter-wave phase at origin since mother-waves phase at origin have been set to zero. 
The steepness is small during this experiment so the phase-locking is visible even on the most distant probes. This phase-locking is a second evidence of the generation of the daughter wave by resonant interactions.

\begin{figure*}
\begin{center}
\psfrag{a}[b][]{$a_4(d) / [ \varepsilon_1^2  \varepsilon_3  d ]$}
\psfrag{t}[t][]{Angle $\theta$ ($^{\rm o}$)}
\psfrag{f}[][]{$\theta_m$}
\psfrag{b}[][]{$r < 1$}
\psfrag{c}[][]{($f_1<f_3$)}
\psfrag{d}[][]{$r > 1$}
\psfrag{e}[][]{($f_1>f_3$)}
\includegraphics[scale=0.38]{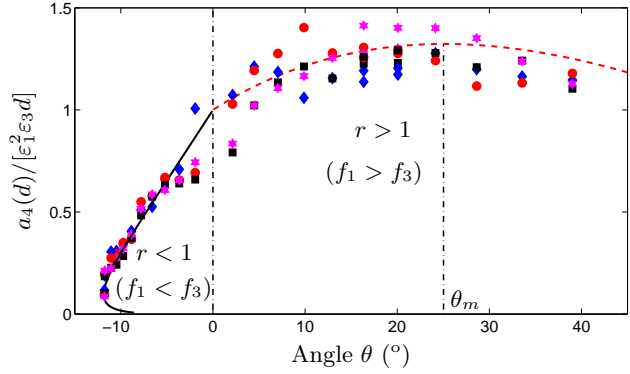}
\caption{Rescaled amplitude $a_4 / (\varepsilon_1^2  \varepsilon_3  d)$ vs. angle $\theta_{}$ for different distances $d=7.8$ ($\blacklozenge$), 9.9 ($\bullet$), 11.9 ($\blacksquare$), and 13.8 ($\ast$) m. Theoretical resonance curve $G[\theta_{}(r)]$ for $r<1$ (solid black line) and $r>1$ (dashed red/gray line) from \citet{LH62}. $\varepsilon_1=\varepsilon_3=0.07$. $f_1=0.9$ Hz. $0.83 \leq r\equiv f_1/f_3 \leq 1.38$. $\theta_{m}=25^{\mathrm o}$. } 
%\todo{Figure  \`a refaire en rempla\c cant les 2 par des 3 et en faisant des labels propres pour les axes (virer les *).}
% d o s h
\label{fig06}
\end{center}
\end{figure*}
%\caption{Rescaled amplitude $a_4$ measured at different distances d for out-of-resonance conditions. Main: $a_4$ vs normalized-centered detuning $r^{\star}\equiv(r-r_m)/(r_0-r_m)$. $r_0$ is the first zero of the sinc function. Inset: rescaled $a_4$ vs normalized detuning $r/r_m$. Symbols corresponds to different $d=7$ up to 27 m (see arrows ). Frequency detuning: $0.96\leq r \leq 1.56$. Resonance: $r_m=1.258$. Solid line: normalized sinc function $|\sin{(\pi dr^{\star})} / (\pi d r^{\star})|$ \citep{LHS66}. $\varepsilon_1=\varepsilon_3=0.07$. $f_1=0.9$ Hz.}
%

The figure-of-eight is now investigated in the vicinity of maximum growth rate (see figure \ref{fig02}, left). In the dedicated experiments B, the mother-wave angle $\theta_{}$ is varied in the range from -10$^{\rm o}$ to 0$^{\rm o}$ in the case $r<1$ (or $f_3>f_1$) and from 0$^{\rm o}$ to +40$^{\rm o}$ in the case $r>1$. For each angle $\theta$, the frequency $f_3$ is chosen so that ${\mathbf k_3}$ is located on the figure-of-eight (see figure \ref{fig02}, right) in order to fulfill the resonance conditions. Note that the correct choice of the directions $\theta_1$ and $\theta_3$ of the individual mother waves in the basin is a key point in obtaining significant results. The successful strategy is to ensure the direction of daughter wave 4 follows the line of the probes. Figure \ref{fig06} shows the rescaled daughter-wave amplitude $a_4/(\varepsilon_1^2\varepsilon_3 d)$ as a function of the angle $\theta_{}$ for different distances $d$ at fixed steepnesses $\varepsilon_1$ and $\varepsilon_3$. This rescaling allows to measure experimentally the resonance response curve $G(\theta)$ predicted by \citet{LH62}. For all values of $\theta$, a good quantitative agreement with the theoretical $G(\theta)$ is observed with no fitting parameter. This strongly extends previous experiments \citep{LHS66,McGoldrick66,tomita_89}, which were carried out only for perpendicular conditions ($\theta_{}=90^{\rm o}$).

\section{Out-of-resonance experiments}

\begin{figure*}
\begin{bigcenter}
\psfrag{a}[b][]{$a_4/(\varepsilon_1^2 \varepsilon_3 d G(r_m))$}
%\psfrag{d}[t][]{$r/r_m$}
\psfrag{d}[t][]{$\Delta k$}
{\includegraphics[scale=0.35]{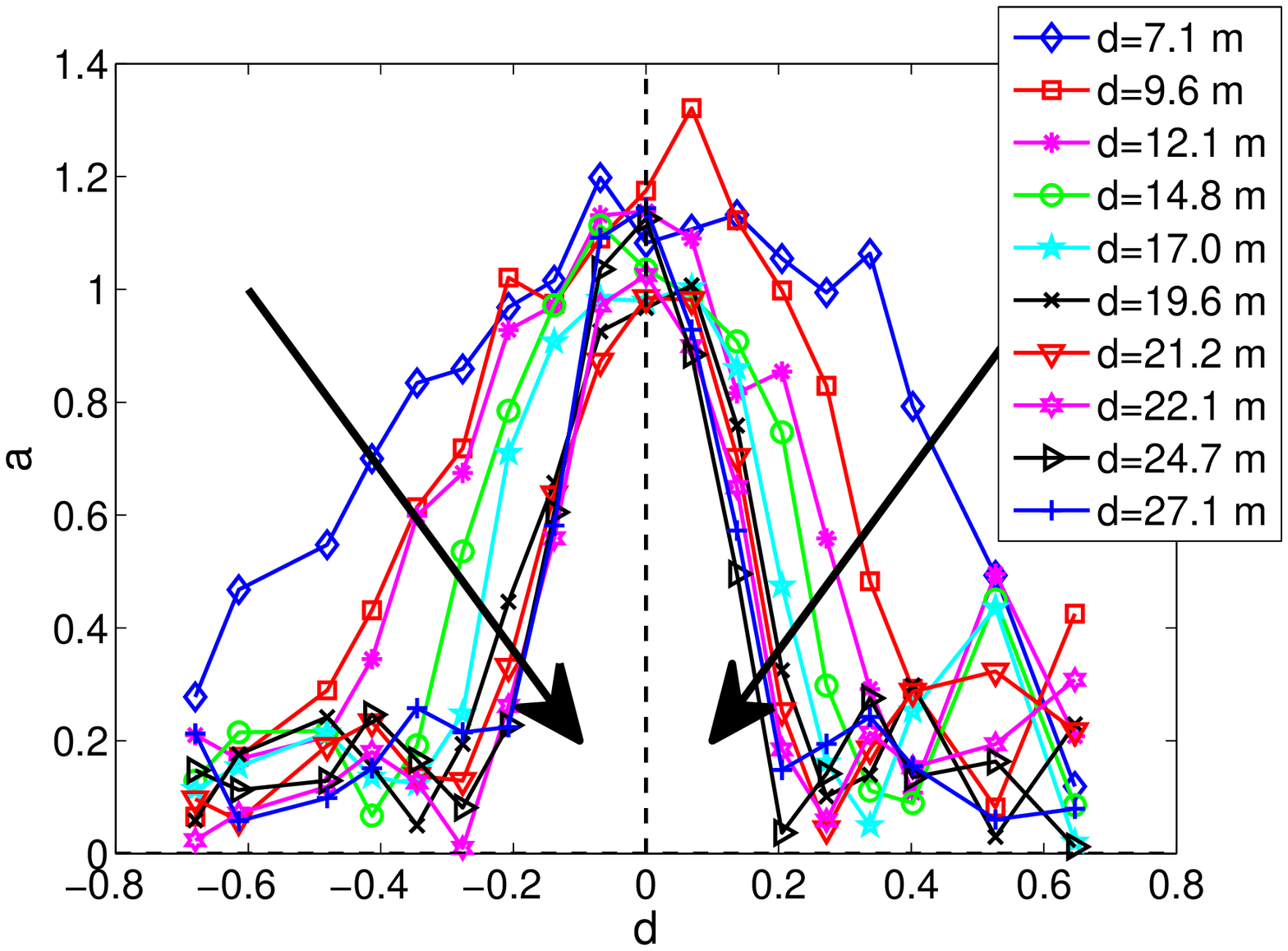}}
\psfrag{b}[b][]{$a_4/(\varepsilon_1^2 \varepsilon_3 d G(r_m))$}
%\psfrag{r}[t][]{$r^*=\frac{r-r_m}{r_0-r_m}$}
\psfrag{d}[t][]{$\Delta k d / 2$}
\psfrag{-3pi}[][]{$-3\pi$}
\psfrag{-2pi}[][]{$-2\pi$}
\psfrag{-pi}[][]{$-\pi$}
\psfrag{pi}[][]{$\pi$}
\psfrag{2pi}[][]{$2\pi$}
\psfrag{3pi}[][]{$3\pi$}
\hspace{5mm}
{\includegraphics[scale=0.35]{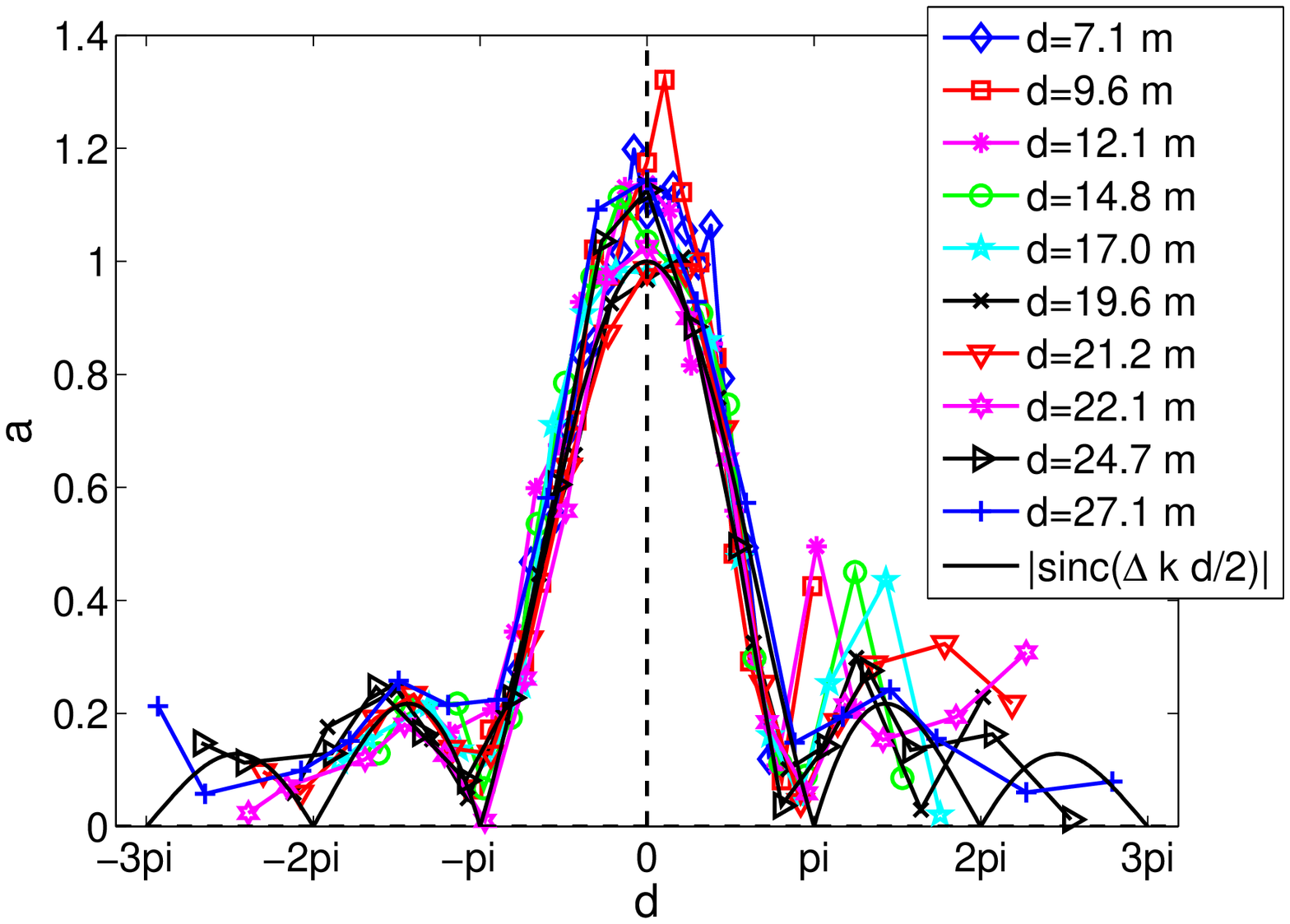}}
%\caption{Rescaled amplitude $a_4/(\varepsilon_1^2 \varepsilon_3 d G(r))$ measured at different distances d for out-of-resonance conditions ($\varepsilon_1=\varepsilon_3=0.07$ and $f_1=0.9$ Hz.). Left: rescaled $a_4$ vs normalized detuning $r/r_m$. Symbols corresponds to different $d=7$ up to 27 m (see arrows). Frequency detuning: $0.96\leq r \leq 1.56$. Resonance: $r_m=1.258$. Solid line: normalized sinc function $|\sin{(\pi r^{\star})} / (\pi r^{\star})|$ \citep{LHS66}. Right: $a_4$ vs normalized-centered detuning $r^{\star}\equiv(r-r_m)/(r_0-r_m)$. $r_0$ is the first zero of the sinc function.}
\caption{Rescaled amplitude $a_4/(\varepsilon_1^2 \varepsilon_3 d G(r_m))$ measured at different distances $d$ for out-of-resonance conditions ($\varepsilon_1=\varepsilon_3=0.07$ and $f_1=0.9$ Hz.). Left: rescaled $a_4$ vs detuning $\Delta k$. Symbols corresponds to different $d=7$ up to 27 m (see arrows). Right: rescaled $a_4$ vs normalized detuning $\Delta k d/2$. Solid line: absolute sinc function $|\mathrm{sinc} \Delta k d/2|$ from \citet{LH62} estimation or from equation \eqref{eq:eps4noreso}. }
\label{fig07}
% info C6Serie1et8
\end{bigcenter}
\end{figure*}

Let us now turn to experiments with out-of-resonance conditions for mechanically generated mother waves. \modif{These conditions} correspond to $2\omega_1-\omega_3-\omega_4=0$ and $2{\mathbf k_1}-{\mathbf k_3} - {\mathbf k_4}\equiv\Delta {\mathbf k}\neq {\mathbf 0}$. Although the direction of the daughter wave is not specified, we assume that the fastest growing daughter wave is the one with minimal detuning. In other words, the daughter wave propagates along the direction of $2{\mathbf k_1}-{\mathbf k_3}$ and the corresponding detuning is now $\Delta k\equiv\left|2{\mathbf k_1}-{\mathbf k_3}\right|-k(2\omega_1-\omega_3)$. %, this linear wavenumber detuning being small $\Delta k/k_3\ll 1$ hereafter. 
We investigate experimentally this case (set C) near the location of the maximum growth rate at $r=r_m$. To wit, we kept the same angle $\theta_{}=\theta_{m}$ and varied the frequency $f_3$ so that ${\mathbf k_3}$ can deviate from the figure of eight (see the green dashed line in Figure \ref{fig02}, right). Figure \ref{fig07}, left, shows the normalized daughter-wave amplitude defined by $a_4/(\varepsilon_1^2 \varepsilon_3 d G(r_m))$ as a function of the detuning $\Delta k$ for different distances $d$ . We observe a decrease of the resonance bandwidth with increasing distance as expected from the $\mathrm{sinc}$ term in equation \eqref{eq:eps4noreso}. 
%experiments with $\varepsilon_1=\varepsilon_3=0.07$
We rescaled all these curves on a single curve as shown on the right in Fig. \ref{fig07} by scaling the detuning with half the distance. We observe that all our measurements collapse on the $\mathrm{sinc}$ curve showing a good agreement with estimation from \citet{LH62} or from equation \eqref{eq:eps4noreso} rigorously derived. 
\section{Conclusion}
We have presented experiments on resonant interactions of surface gravity waves within the Ecole Centrale de Nantes wave basin (50 m long by 30 m large by 5 m deep) in a degenerated case. Bichromatic mother waves were generated mechanically by means of specific control of oblique wave generation (Dalrymple method). The linear spatial growth of a resonant daughter wave was observed. The theoretical and experimental results presented here extend the pioneering work done in the 60s. Four-wave interaction theory is expressed in the framework of Hamiltonian dynamic theory to demonstrate a phase-locking mechanism for resonant quartets and estimate the daughter-wave amplitude in nearly-resonant quartets. All these theoretical results are supported by experimental observations of generated oblique mother waves: the observed linear spatial growth-rate of daughter wave scaling with mother-wave steepness; the phase-locking between resonant waves; the growth rate $G$ satisfying the law historically found by \citet{LH62}; as well as the off-resonance response following the expected $\mathrm{sinc}$ curve.

The experiments presented in this article correspond to the early stage of resonance, that is when $k_4 \varepsilon^2 d <1$. Indeed, for longer distance or greater steepness, we observed other common features of nonlinear interactions at resonance (not reported in this paper) such as the pumping of the mother wave by the resonant wave and the decrease of resonant wave growth. For off-resonance conditions and stronger wave steepness ($ka > 0.1$), departures from the approximate off-resonance equation \eqref{eq:eps4noreso} are observed: distortion of the response curve ($\mathrm{sinc}$) by a nonlinear detuning. % (when $\Delta_k/k_3$ is not small). 
These nonlinear effects will be the subject of a further publication. The Hamiltonian theory may serve as an extension of the theory in \citet{LH62} to higher steepness, either by analytical solutions (see {\rm e.g.} \citet{Stiassnie2005}) or numerical solutions \citep{Leblanc2009}. Finally, experiments with much greater steepness should allow quantification of the departure from weakly nonlinear theory (Zakharov equation). It would also provide a better understanding of wave turbulence experiments in strongly nonlinear regimes.
\modif{
\begin{acknowledgments}
This work was supported by ANR Turbulon 12-BS04-0005. We thank C. Laroche and A. Levesque for their technical help. We also thank H. Houtani and T. Waseda for providing us the report \citet{tomita_89}.
\end{acknowledgments}
}
\bibliographystyle{jfm}
\bibliography{LH}

\begin{thebibliography}{26}
\expandafter\ifx\csname natexlab\endcsname\relax\def\natexlab#1{#1}\fi

\bibitem[Aubourg \& Mordant(2015)]{aubourg2015}
{\sc Aubourg, Q. \& Mordant, N.} 2015 Nonlocal resonances in weak turbulence of
  gravity-capillary waves. {\em Phys. Rev. Lett.\/} {\bf 114}, 144501.

\bibitem[Bonnefoy {\em et~al.\/}(2015)Bonnefoy, Haudin, Michel, Semin, Humbert,
  Auma{\^i}tre, Berhanu \& Falcon]{theory}
{\sc Bonnefoy, F., Haudin, F., Michel, G., Semin, B., Humbert, T.,
  Auma{\^i}tre, S., Berhanu, M. \& Falcon, E.} 2015 Other supplementary
  material, observation of resonant interactions among gravity surface waves.

\bibitem[Bordes {\em et~al.\/}(2012)Bordes, Moisy, Dauxois \& Cortet]{BMDC12}
{\sc Bordes, G., Moisy, F., Dauxois, T. \& Cortet, P.-P.} 2012 Experimental
  evidence of a triadic resonance of plane inertial waves in a rotating fluid.
  {\em Phys. of Fluids\/} {\bf 24}~(1).

\bibitem[Dalrymple(1989)]{dalrymple-method-89}
{\sc Dalrymple, R.~A.} 1989 Directional wavemaker theory with sidewall
  reflection. {\em J. Hydr. Res.\/} {\bf 27}~(1), 23--24.

\bibitem[Hammack {\em et~al.\/}(2005)Hammack, Henderson \& Segur]{hammack05}
{\sc Hammack, J.~L., Henderson, D.~M. \& Segur, H.} 2005 Progressive waves with
  persistent two-dimensional surface patterns in deep water. {\em J. Fluid
  Mech.\/} {\bf 532}, 1--52.

\bibitem[Haudin {\em et~al.\/}(2016)Haudin, Cazaubiel, Deike, Jamin, Falcon \&
  Berhanu]{haudin_16}
{\sc Haudin, F., Cazaubiel, A., Deike, L., Jamin, T., Falcon, E. \& Berhanu,
  M.} 2016 Experimental study of three-wave interactions among
  capillary-gravity surface waves. {\em Phys. Rev. E\/} {\bf 93}, 043110.

\bibitem[Henderson \& Hammack(1987)]{HH87}
{\sc Henderson, D.~M. \& Hammack, J.~L.} 1987 Experiments on ripple
  instabilities. part 1. resonant triads. {\em J. Fluid Mech.\/} {\bf 184},
  15--41.

\bibitem[Janssen(2009)]{janssen09}
{\sc Janssen, P. A. E.~M.} 2009 On some consequences of the canonical
  transformation in the hamiltonian theory of water waves. {\em J. Fluid
  Mech.\/} {\bf 637}, 1--44.

\bibitem[Joubaud {\em et~al.\/}(2012)Joubaud, Munroe, Odier \& Dauxois]{JMOD12}
{\sc Joubaud, S., Munroe, J., Odier, P. \& Dauxois, T.} 2012 Experimental
  parametric subharmonic instability in stratified fluids. {\em Phys. of
  Fluids\/} {\bf 24}~(4).

\bibitem[Krasitskii(1994)]{krasitskii94}
{\sc Krasitskii, V.~P.} 1994 On reduced equations in the hamiltonian theory of
  weakly nonlinear surface waves. {\em J. Fluid Mech.\/} {\bf 272}, 1 -- 20.

\bibitem[Lake \& Yuen(1977)]{lake77}
{\sc Lake, B.M. \& Yuen, H.C.} 1977 A note on some nonlinear water-wave
  experiments and the comparison of data with theory. {\em J. Fluid Mech.\/}
  {\bf 83}, 75--81.

\bibitem[Leblanc(2009)]{Leblanc2009}
{\sc Leblanc, S.} 2009 Stability of bichromatic gravity waves on deep water.
  {\em Eur. J. Mech. / B Fluids\/} {\bf 28}~(5), 605--612.

\bibitem[Liu {\em et~al.\/}(2015)Liu, Xu, Li, Peng, Alsaedi \& Liao]{Liu2015}
{\sc Liu, Z., Xu, D.~L., Li, J., Peng, T., Alsaedi, A. \& Liao, S.~J.} 2015 On
  the existence of steady-state resonant waves in experiments. {\em J. Fluid
  Mech.\/} {\bf 763}, 1--23.

\bibitem[Longuet-Higgins(1962)]{LH62}
{\sc Longuet-Higgins, M.~S.} 1962 Resonant interactions between two trains of
  gravity waves. {\em J. Fluid Mech.\/} {\bf 12}, 321--32, \\{W}e have noticed
  a misprint in equation (6.4) in \citet{LH62}: the term $-(6+\xi^2)^{1/2}$
  should be replaced by $-{\rm sgn}(\xi)(6+\xi^2)^{1/2}$ where $\xi=(1-r)/r$.

\bibitem[Longuet-Higgins \& Smith(1966)]{LHS66}
{\sc Longuet-Higgins, M.~S. \& Smith, N.~D.} 1966 An experiment on third-order
  resonant wave interactions. {\em J. Fluid Mech.\/} {\bf 25}, 417--435.

\bibitem[Martin {\em et~al.\/}(1972)Martin, Simmons \& Wunsch]{MSW72}
{\sc Martin, B.~S., Simmons, W. \& Wunsch, C.} 1972 The excitation of resonant
  triads by single internal waves. {\em J. Fluid Mech.\/} {\bf 53}, 17--44.

\bibitem[McGoldrick(1970)]{MCG70b}
{\sc McGoldrick, L.~F.} 1970 An experiment on second-order capillary gravity
  resonant wave interactions. {\em J. Fluid Mech.\/} {\bf 40}, 251--271.

\bibitem[McGoldrick {\em et~al.\/}(1966)McGoldrick, Phillips, Huang \&
  Hodgson]{McGoldrick66}
{\sc McGoldrick, L.~F., Phillips, O.~M., Huang, N.~E. \& Hodgson, T.~H.} 1966
  Measurements of third-order resonant wave interactions. {\em J. Fluid
  Mech.\/} {\bf 25}, 437--456.

\bibitem[Phillips(1960)]{PhillipsJFM60}
{\sc Phillips, O.~M.} 1960 On the dynamics of unsteady gravity waves of finite
  amplitude. part {I}. the elementary interactions. {\em J. Fluid Mech.\/} {\bf
  9}, 193--217.

\bibitem[Shemer \& Chamesse(1999)]{shemer99}
{\sc Shemer, L. \& Chamesse, M.} 1999 Experiments on nonlinear
  gravity–capillary waves. {\em J. Fluid Mech.\/} {\bf 380}, 205--232.

\bibitem[Stiassnie \& Shemer(2005)]{Stiassnie2005}
{\sc Stiassnie, M. \& Shemer, L.} 2005 On the interaction of four water-waves.
  {\em Wave Motion\/} {\bf 41}~(4), 307 -- 328.

\bibitem[Su {\em et~al.\/}(1982)Su, Bergin, Marler \& Myrick]{su82}
{\sc Su, M.-Y., Bergin, M., Marler, P. \& Myrick, R.} 1982 Experiments on
  nonlinear instabilities and evolution of steep gravity-wave trains. {\em J.
  Fluid Mech.\/} {\bf 124}, 45--72.

\bibitem[Tomita(1989)]{tomita_89}
{\sc Tomita, H.} 1989 Theoretical and experimental investigations of
  interaction among deep-water gravity waves. {\em Report of Ship Res. Inst.\/}
  {\bf 26}~(5), 251--350.

\bibitem[Tulin \& Waseda(1999)]{tulin-waseda-JFM99}
{\sc Tulin, M.~P. \& Waseda, T.} 1999 Laboratory observations of wave group
  evolution, including breaking effects. {\em J. Fluid Mech.\/} {\bf 378},
  197--232.

\bibitem[Waseda {\em et~al.\/}(2015)Waseda, Kinoshita, Cavaleri \&
  Toffoli]{waseda15}
{\sc Waseda, T., Kinoshita, T., Cavaleri, L. \& Toffoli, A.} 2015 Third-order
  resonant wave interactions under the influence of background current fields.
  {\em J. Fluid Mech.\/} {\bf 784}, 51--73.

\bibitem[Zakharov(1968)]{Zakharov68}
{\sc Zakharov, V.} 1968 Stability of periodic waves of finite amplitude on a
  surface of a deep fluid. {\em J. Appl. Mech. Tech. Phys.\/} {\bf 2},
  190--198.

\end{thebibliography}


\begin{thebibliography}{12}
\expandafter\ifx\csname natexlab\endcsname\relax\def\natexlab#1{#1}\fi

\bibitem[Boyd(2008)]{Boyd08}
{\sc Boyd, R.~W.} 2008 {\em Nonlinear Optics, Third Edition\/}, 3rd edn.
  Academic Press.

\bibitem[Hudspeth \& Sulisz(1991)]{hudspeth-stokes-91}
{\sc Hudspeth, R.~T. \& Sulisz, W.} 1991 Stokes drift in two-dimensional wave
  flumes. {\em J. Fluid Mech.\/} {\bf 230}, 209--229.

\bibitem[Janssen(2009)]{janssen09}
{\sc Janssen, P. A. E.~M.} 2009 On some consequences of the canonical
  transformation in the hamiltonian theory of water waves. {\em J. Fluid
  Mech.\/} {\bf 637}, 1--44.

\bibitem[Krasitskii(1994)]{krasitskii94}
{\sc Krasitskii, V.~P.} 1994 On reduced equations in the hamiltonian theory of
  weakly nonlinear surface waves. {\em J. Fluid Mech.\/} {\bf 272}, 1 -- 20.

\bibitem[Leblanc(2009)]{Leblanc2009}
{\sc Leblanc, S.} 2009 Stability of bichromatic gravity waves on deep water.
  {\em Eur. J. Mech. / B Fluids\/} {\bf 28}~(5), 605--612.

\bibitem[Longuet-Higgins(1962)]{LH62}
{\sc Longuet-Higgins, M.~S.} 1962 Resonant interactions between two trains of
  gravity waves. {\em J. Fluid Mech.\/} {\bf 12}, 321--32, \\{W}e have noticed
  a misprint in equation (6.4) in \citet{LH62}: the term $-(6+\xi^2)^{1/2}$
  should be replaced by $-{\rm sgn}(\xi)(6+\xi^2)^{1/2}$ where $\xi=(1-r)/r$.

\bibitem[Longuet-Higgins \& Smith(1966)]{LHS66}
{\sc Longuet-Higgins, M.~S. \& Smith, N.~D.} 1966 An experiment on third-order
  resonant wave interactions. {\em J. Fluid Mech.\/} {\bf 25}, 417--435.

\bibitem[McGoldrick {\em et~al.\/}(1966)McGoldrick, Phillips, Huang \&
  Hodgson]{McGoldrick66}
{\sc McGoldrick, L.~F., Phillips, O.~M., Huang, N.~E. \& Hodgson, T.~H.} 1966
  Measurements of third-order resonant wave interactions. {\em J. Fluid
  Mech.\/} {\bf 25}, 437--456.

\bibitem[Stiassnie \& Shemer(2005)]{Stiassnie2005}
{\sc Stiassnie, M. \& Shemer, L.} 2005 On the interaction of four water-waves.
  {\em Wave Motion\/} {\bf 41}~(4), 307 -- 328.

\bibitem[Tomita(1989)]{tomita_89}
{\sc Tomita, H.} 1989 Theoretical and experimental investigations of
  interaction among deep-water gravity waves. {\em Report of Ship Res. Inst.\/}
  {\bf 26}~(5), 251--350.

\bibitem[Zakharov(1968)]{Zakharov68}
{\sc Zakharov, V.} 1968 Stability of periodic waves of finite amplitude on a
  surface of a deep fluid. {\em J. Appl. Mech. Tech. Phys.\/} {\bf 2},
  190--198.

\bibitem[Zakharov {\em et~al.\/}(1992)Zakharov, Lʹvov \&
  Falkovich]{zakharov1992kolmogorov}
{\sc Zakharov, V.E., Lʹvov, V.S. \& Falkovich, G.} 1992 {\em Kolmogorov
  spectra of turbulence\/}. {\em Springer series in nonlinear dynamics\/}
  vol.~1. Springer-Verlag.

\end{thebibliography}

\end{document}

\maketitle

\section{Introduction}
In this supplementary material, we derive the equations that are used in the main article on  degenerated resonance. Hamiltonian theory is presented following \citet{Zakharov68}. Such a model yields slow-time evolution of complex wave amplitudes and receives continuous interest, especially in four-wave interactions (see e.g. \citet{Stiassnie2005,janssen09,Leblanc2009}).

We use this Hamiltonian approach to derive a solution at short-time in degenerated four-wave interactions in order to explain the $\rm{sinc}$ detuning (or phase mismatch) factor first introduced in \citet{LHS66} when the waves are off-resonance. \citet{LHS66} based an explanation for this $\mathrm{sinc}$ term on the superposition of the near-resonant daughter wave and a free wave generated by the wavemaker. Both waves have the same amplitude to ensure a zero-flux boundary condition. Saying this, they neglected the evanescent waves which are known to make an important contribution to free wave generation \citep{hudspeth-stokes-91}. 

Assuming constant mother-wave amplitudes (see \citet{Boyd08} in optics), we recover all the previous results in \citet{LH62}, including the phase-locking observed also in our experiments. We show that for resonance or near-resonance, the total phase is found to be initially locked to $\pi/2$ (valid only for short times) and then to slowly evolve away from this initial value. Concerning the phase mismatch factor, note that \citet{tomita_89} already used the same Hamiltonian derivation and found the intermediate sine solution for the daughter-wave amplitude yet without linking it to the detuning $\rm{sinc}$ behavior first described by \citet{LH62}. In order to improve the accessibility of this type of results, we present the relevant theory in readable form with full details in this supplementary material.
\section{Approximate Hamiltonian Theory}\label{sec:theory}
\subsection{General case}
The dynamical Hamiltonian theory is presented here to account for phase evolution and off-resonance solution; we follow the formalism from \citet{janssen09} and \citet{zakharov1992kolmogorov}. The potential flow unknowns are the free surface elevation $\eta({\mathbf x},t)$ and the free surface potential $\psi({\mathbf x},t)$. The latter is the value of the potential of the flow $\phi$ taken at the free surface, that is $\psi({\mathbf x},t)=\phi({\mathbf x},z=\eta({\mathbf x},t),t)$. The corresponding space Fourier transforms $\hat{\eta}$ and $\hat{\psi}$ are defined by
\begin{eqnarray}
\hat{\eta}({\mathbf k},t) &=& \frac{1}{2\pi} \, \int d{\mathbf x} \, \eta({\mathbf x},t) \, \exp\left( - \ic {\mathbf k}.{\mathbf x} \right) {\rm \ ,}
\\
\hat{\psi}({\mathbf k},t) &=& \frac{1}{2\pi} \, \int d{\mathbf x} \, \psi({\mathbf x},t)\, \exp\left( - \ic {\mathbf k}.{\mathbf x} \right) {\rm \ .}
\label{eq:Fourier}
\end{eqnarray}
For each wavevector ${\mathbf k}$, the corresponding frequency $\omega$ is given by the considered dispersion relation $\omega({\mathbf k})$. These transforms are multiplied by $\sqrt{\omega/k}$ and $\sqrt{k/\omega}$ respectively where $k=|{\mathbf k}|$ so that the resulting amplitudes have the same dimension. After this first canonical transformation\footnote{By definition, a canonical transformation preserves the form of the Hamilton's equations describing the system.}, the complex action variable $A({\mathbf k})$ is defined as follows by a second canonical transformation
\begin{equation}
A({\mathbf k},t) \egal \frac{1}{2^{1/2}} \, \left[\left( \frac{\omega}{k} \right)^{1/2}\hat{\eta}({\mathbf k},t)  + \ic \left(\frac{k}{\omega} \right)^{1/2}\hat{\psi}({\mathbf k},t)\right]{\rm \ .}
\label{eq:cano1}
\end{equation}
We may use later on the following relations
\begin{eqnarray}
\hat{\eta}({\mathbf k},t) & = & \left( \frac{k}{2\omega} \right)^{1/2} \left[ A({\mathbf k},t) + A^*(-{\mathbf k},t)\right]{\rm \ ,}\\
\hat{\psi}({\mathbf k},t) & = & -\ic \left( \frac{\omega}{2k} \right)^{1/2} \left[ A({\mathbf k},t) - A^*(-{\mathbf k},t)\right]{\rm \ .}
\label{eq:cano2}
\end{eqnarray}

A third canonical transformation from $A({\mathbf k},t)$ to a new variable $\hat{a}({\mathbf k},t)$ eliminates the non-resonant interactions describing bound waves in the three-wave and four-wave processes. The reader will refer to \citet{janssen09} and \citet{zakharov1992kolmogorov} for more details. By acknowledging the linear evolution of the action variable, a new unknown variable is introduced $B({\mathbf k},t)=\hat{a}({\mathbf k},t) \exp (\ic \omega(k) t)$ called action amplitude or generalized amplitude spectrum. %In $B$ the linear evolution of the phase is already taken into account. 
Further equations for $B$ assess only the nonlinear part. For small wave steepness, the nonlinear evolution of waves with non-decay relation dispersion\footnote{All the three-wave interactions are therefore non-resonant.} ($\omega \propto k^\nu$ with $\nu<1$) is described by Zakharov's equation \citep{Zakharov68}
\begin{equation}
i \partial_t B_1 = \iiint T_{1234} B_2^* B_3 B_4 \delta_{1+2-3-4} \, \exp\left(\ic\Delta_{1234}t\right) \, d {\mathbf k_2}d{\mathbf k_3}d{\mathbf k_4}{\rm \ ,}
\label{eq:zakharov}
\end{equation}
where the frequencies are $\omega_i=\omega({\mathbf k_i})$, the frequency detuning or mismatch is $\Delta_{1234}=\omega_1+\omega_2-\omega_3-\omega_4$, $\delta_{1+2-3-4}=\delta({\mathbf k_1}+{\mathbf k_2}-{\mathbf k_3}-{\mathbf k_4})$ and $B_i=B({\mathbf k_i},t)$ the notation for the action. The interaction coefficients $T_{1234} = T({\mathbf k_1},{\mathbf k_2},{\mathbf k_3},{\mathbf k_4})$ are the kernels given in \citet{krasitskii94} or \citet{janssen09}. 

The free-surface elevation is related to the generalized amplitude spectrum $B({\mathbf k},t)$ by the above canonical transformations. The nonlinear elevation consists of a linear superposition of free waves and a ensemble of corresponding bound waves. The linear part of the elevation is build as the superposition of free waves 
\begin{equation}
\eta_{lin}({\mathbf x},t) = \int d{\mathbf k} \left( \frac{k}{2\omega} \right)^{1/2} \left[ B({\mathbf k},t)\exp (-\ic \omega(k) t) + B^*(-{\mathbf k},t)\exp (\ic \omega(k) t)\right] \, \exp\left( \ic {\mathbf k}.{\mathbf x} \right) {\rm \ .}
\label{eq:eta}
\end{equation}
The bound waves can also be computed by means of the canonical transformations (see \citet{janssen09} for the corresponding kernels).
\subsection{Degenerated resonance}
The degenerated case we study in the paper consists of only three waves, two mother waves 1 and 3 and a daughter wave 4. The wave action amplitude is noted $B({\mathbf k},t)=B_1(t)\delta({\mathbf k}-{\mathbf k}_1)+B_3(t)\delta({\mathbf k}-{\mathbf k}_3)+B_4(t)\delta({\mathbf k}-{\mathbf k}_4)$ with the resonance condition $2{\mathbf k_1} - {\mathbf k_3} - {\mathbf k_4}={\mathbf 0}$. In the case of an homogeneous wave field, the equations for the degenerated case are deduced from equation \eqref{eq:zakharov} 
\begin{subequations}\label{eq:zakh-degen}
\begin{align}
\ic\partial_t B_1 &= (\Omega_1-\omega_1) B_1 + 2T_{1134} \exp(\ic\Delta\omega t) B_1^* B_3 B_4 {\rm \ ,}
\label{eq:B1}\\
\ic\partial_t B_3 &= (\Omega_3-\omega_3) B_3 + \,\,\,T_{1134} \exp(-\ic\Delta\omega t) B_1^2 B_4^* {\rm \ ,}
\label{eq:B3}\\
\ic\partial_t B_4 &= (\Omega_4-\omega_4) B_4 + \,\,\,T_{1134} \exp(-\ic\Delta\omega t) B_1^2 B_3^* {\rm \ ,}
\label{eq:B4}
\end{align}
\end{subequations}
where $\Delta\omega=2\omega_1-\omega_3-\omega_4$ is the linear frequency detuning. Nonlinear frequencies $\Omega_i$ satisfy the following nonlinear dispersion relations 
\begin{equation}
\left.\begin{array}{rcl}
\Omega_1 &=& \omega_1+\,\,\,T_{1111} |B_1|^2+2T_{1313} |B_3|^2+2T_{1414} |B_4|^2 {\rm \ ,}
\\
\Omega_3 &=& \omega_3+2T_{1313} |B_1|^2+\,\,\,T_{3333} |B_3|^2+2T_{3434} |B_4|^2 {\rm \ ,}
\\
\Omega_4 &=& \omega_4+2T_{1414} |B_1|^2+2T_{3434} |B_3|^2+\,\,\,T_{4444} |B_4|^2 {\rm \ .}
\end{array}\right\}
\label{eq:NL-disp}
\end{equation}
Here we consider degenerated wave fields where only the two mother waves 1 and 3 are initially present, \ie\ $B_4(t=0)=0$. In this case, the resonant daughter wave exhibits linear growth in the early stage of the four-wave interaction and an energy exchange happens from mother wave 1 towards mother wave 3 and daughter wave 4. Equations \eqref{eq:zakh-degen} admit self-similar solutions of the form $B_i=B_{i0} \, f_i(\alpha^2 t,\beta)$ for $i=1,3$ with $B_{i0}$ the initial amplitude and $f_i$ functions of unit magnitude; index 0 denotes the solutions at $t=0$. The daughter-wave amplitude may also be written as $B_4=\alpha \, f_4(\alpha^2t,\beta)$ where $\alpha$ is the scale $|B_{10}^2 B_{30}|^{1/3}$ and $\beta=|B_{30}/B_{10}|$. Such solutions may be obtained analytically (see \citet{Stiassnie2005}) or numerically. 

In the following, we define the total detuning $\Delta\Omega = 2\Omega_1-\Omega_3-\Omega_4$ and the detuning due to nonlinear effects $\Delta \omega_{nl}=\Delta\Omega-\Delta\omega$. 
\subsection{Solution at small daughter-wave amplitude}
In the early stage of the resonant interaction or for a non-resonant interaction, the daughter-wave amplitude is assumed to be negligible with respect to the mother-wave amplitudes. In equations \eqref{eq:B1} and \eqref{eq:B3}, the first term of the right-hand side is dominant and it follows that the mother waves evolve solely due to the nonlinear dispersion; in other words, they keep a constant amplitude $B_{i0}$ for $i=1$ and 3. In equations \eqref{eq:NL-disp}, the third term of the right-hand side disappear and the mother-wave nonlinear frequency $\Omega_i$ as well as the daughter-wave one $\Omega_4$ are also constant.

Reporting this in equations \eqref{eq:B1} and \eqref{eq:B3}, we obtain the mother-wave amplitudes
\begin{equation}
B_i(t)=B_{i0} \, \exp(-\ic(\Omega_i-\omega_i)t) \, , \,\,i=1  \, {\rm and } \, 3{\rm \ .}
\label{eq:B1-B3}
\end{equation}
Introducing $C_4$ such as $B_4(t)=C_4(t) \, \exp(-\ic(\Omega_4-\omega_4)t)$, we obtain from equation \eqref{eq:B4} a new equation for $C_4$
\begin{equation}
\partial_t C_4 = -\ic\, T_{1134} \, \exp(-\ic\Delta\omega t) B_{10}^2 B_{30}^* \, \exp\left(-\ic \Delta \omega_{nl} t\right) {\rm \ .}
\label{eq:C4}
\end{equation}
We can see that amplitude $C_4$ accounts for energy transfer as well as for all nonlinear frequency evolution due to the interaction other than the nonlinear dispersion. We have now after straightforward integration of equation \eqref{eq:C4}
\begin{equation}
B_4 = - \ic \, T_{1134}  B_{10}^2 B_{30}^* \, \frac{\sin\left(\Delta \Omega t/2\right)}{\Delta \Omega/2}\, \exp(-\ic(\Omega_4-\omega_4+\Delta \Omega/2)t) {\rm \ .}
\label{eq:B4-sol}
\end{equation}
This solution \eqref{eq:B4-sol} is valid as long as $|B_4|\ll |B_{10}|$ and $|B_{30}|$. This expression of the complex amplitude provides the daughter-wave real amplitude and phase. First, since $T_{1134}>0$ (see \cite{janssen09}), the daughter-wave amplitude is
\begin{equation}
|B_4| = T_{1134} \, | B_{10}^2 B_{30} | \, \left|\frac{\sin\left(\Delta \Omega t/2\right)}{\Delta \Omega/2} \right| {\rm \ .}
\label{eq:B4-mag-sol}
\end{equation}

The linear free surface elevation consists in the surperposition of three waves
$$
\eta_{lin}=\frac{1}{2}\left(\sum_i a_i \exp(\ic ({\mathbf k_i}.{\mathbf x}-\omega_i t+\varphi_i(t))) +c.c. \right){\rm \ ,}
$$
where $i$=1, 3 and 4, $a_i$ is the wave amplitude and $\varphi_i(t)$ is the phase due to nonlinear effects, which evolves slowly in time.
Using equation \eqref{eq:eta} and the previous solution for $B_i(t)$, we obtain the mother-wave phases $\varphi_i(t)=-(\Omega_i-\omega_i)\,t+\varphi_{i0}$ for $i=1$ and 3 where index 0 again denotes initial values. They evolve slowly due to the nonlinear correction in the dispersion relations given in equations \eqref{eq:NL-disp}, whereas the daughter-wave phase taken from equation \eqref{eq:B4-sol} evolves due to two terms
\begin{equation}
\varphi_4(t)= - (\Omega_4-\omega_4)t - \Delta \Omega t/2 + 2\varphi_{10} - \varphi_{30} - \frac{\pi}{2}{\rm \ .}
\label{eq:Phi4}
\end{equation}
The first term is the slow evolution due to nonlinear dispersion correction and the second term comes from the total frequency detuning $\Delta \Omega=\Delta\omega+\Delta \omega_{nl}$ which contains a fast and a slow terms. The last terms show the phase-locking of the daughter wave. Note that the phase $\varphi_4(t)$ could also be defined by $\varphi_4(t)=- (\Omega_4-\omega_4)t - \Delta \Omega t/2 + \varphi_{40}$. We now introduce the total phase $\varphi$ as follows
\begin{equation}
\varphi(t)\equiv 2\varphi_1-\varphi_3-\varphi_4=\frac{\pi}{2} -\frac{\Delta\omega + \Delta \omega_{nl} }{2} t {\rm \ .}
\label{eq:Phi}
\end{equation}
%\frac{\Delta\omega + \Delta }{2} t + 2\varphi_{10}-\varphi_{30}-\varphi_{40}
Both the linear and the nonlinear part of the frequency detuning play a role in the evolution of the total phase which then vary with a fast term $\Delta\omega t/2$ and a slow term $\Delta \omega_{nl} t/2$ on a $\alpha^2 t$ time scale.

The following relation holds between the generalized amplitude spectrum $B_i$ for wave $i$ and the free surface amplitude $\eta_i$
\begin{equation}
a_i = \sqrt{\frac{2k_i}{\omega_i}} \, B_i {\rm \ .}
\label{eq:B2eta}
\end{equation}
It provides us the daughter-wave amplitude $a_4$ as follows
\begin{subequations}\label{eq:a4-mag-phase}
\begin{align}
|a_4| &= T_{1134} \frac{\omega_1}{2k_1^3} \sqrt{\frac{\omega_3 \, k_4}{\omega_4 \, k_3^3}} \varepsilon_{1}^2 \varepsilon_{3} \, \left|\frac{\sin\left(\Delta \Omega t/2\right)}{\Delta \Omega/2}\right|{\rm \ ,}
\label{eq:a4-mag}\\
\arg a_4 &= -\frac{\pi}{2} + 2\arg a_{10}-\arg a_{30} - (\Omega_4-\omega_4+\Delta \Omega/2)t{\rm \ ,}
\label{eq:a4-phase}
\end{align}
\end{subequations}
where the steepness is given by $\varepsilon_i=k_i|a_{i0}|$.
\subsubsection{Exact linear resonance}
At resonance ($\Delta\omega=0$) and for the initial stage of the interaction ($\Delta \omega_{nl} t\ll 1$), equation \eqref{eq:Phi} gives $\varphi=\pi/2$ and equation \eqref{eq:B4-mag-sol} predicts a linear growth of the daughter wave, with a maximum growth rate 
\begin{equation}
|B_4| = T_{1134} \, |B_{10}^2 B_{30}| \, t {\rm \ .}
\label{eq:B4-reso}
\end{equation}
From equation \eqref{eq:Phi4}, phase-locking is expected for the daughter wave whose initial phase is naturally set to $\varphi_{40}=-\pi/2+2\varphi_{10}-\varphi_{30}$. These are the results obtained by means of perturbation theory in \citet{LH62}. Although the phase locking was not explicitly mentionned, it was implicitly accounted for. In \citet{LH62}, the mother-wave profile is described by $\eta_i=a_i\cos \psi_i$ with $\psi_i={\mathbf k_i}.{\mathbf x}-\omega_i t$ for $i=1$ and 3. The evolution of the resonant daughter wave is given by $\eta_4=a_4 \sin (2\psi_1-\psi_3)$. It follows using the resonance conditions that $\sin(2\psi_1-\psi_3)=\cos(\psi_4-\pi/2)$ and hence the phase of the daughter wave is locked to $-\pi/2$ by comparison to the mother waves. 

The bound waves associated with the quartet are given at frequency $2\omega_1-\omega_3$ by $B^{(2)}_{1134} B_{10}^2 B_{30}^*$ (see e.g. \citet{janssen09} for an expression of kernel $B^{(2)}_{1234}$). We have checked numerically (not shown here) that they have negligible amplitudes compared to the resonant daughter wave at the same frequency. 

Concerning the phase evolution, the above solutions for $\varphi_i$ show that all waves phases will evolve with slow nonlinear time (as long as the daughter-wave amplitude is small). From equation \eqref{eq:Phi}, we see that the total phase $\varphi=\pi/2 + \Delta \omega_{nl} t/2$ will evolve from its initial $\pi/2$ value on the long time scale $\alpha^2 t$. In other words, the concept of linear resonance ($\Delta\omega=0$) is valid in the early stage but it does not make sense at longer time since the total phase follow a slow nonlinear evolution. 
%At resonance, the assumption made on  will not hold for long time.
%
\subsubsection{Off-resonance}
We consider now an off-resonance degenerated quartet with a linear frequency detuning $\Delta\omega\neq 0$. Equation \eqref{eq:B4-mag-sol} gives the expression of the off-resonance daughter-wave amplitude, valid when $|B_4|\ll |B_{10}|$ and $|B_{30}|$. For interpretation and following \citet{LH62}, we rewrite equation \eqref{eq:B4-mag-sol} as
\begin{equation}
\frac{|B_4|}{\alpha} = T_{1134} \alpha^2 t \left|\frac{\sin \frac{1}{2}\Delta\Omega t}{\frac{1}{2}\Delta\Omega t} \right| =T_{1134} \alpha^2 t \, \, \left| {\rm sinc} \left( \frac{1}{2}\Delta\Omega t \right) \right| {\rm \ .}
\label{eq:B4-off-sinc}
\end{equation}
In this equation \eqref{eq:B4-off-sinc} we have emphasized on 
\begin{itemize}
	\item the scaling $\alpha=\left|B_{10}^2 B_{30}\right|^{1/3}$ of the daughter-wave amplitude,
	\item the resonant growth $T_{1134} \alpha^2 t$, linear in the slow time scale $\alpha^2 t$,
	\item the off-resonance correction factor ${\rm sinc} \frac{1}{2}\Delta\Omega t$ (known as phase mismatch factor in optics).
\end{itemize}
This last amplitude modulation should not shadow the real evolution observed in equation \eqref{eq:B4-mag-sol} which is a sine function. 
\section{Discussion}
Here the Zakharov equation is applied in the context of degenerated resonant interaction of four waves. The strong approximation we made is to restrict the wave spectrum to three interacting waves only, namely two mother waves present at the start and a daughter wave growing in time. The theoretical developments are not limited to surface gravity waves, and should apply to any nonlinear wave system having forbidden three-wave interactions. 

The solution is found for both resonant and non-resonant cases when the daughter-wave amplitude is small and the solution agrees well with the ones in \citet{LH62} in terms of linear growth rate and in \citet{LHS66} in terms of ${\rm sinc}$ behavior. Note that \citet{LHS66} used a different explanation for the $\rm sinc$ term, based on wavemaker free wave emission, which now seems uncorrect in the light of the dynamical theory used here. In their use of zero-flux boundary condition on the wavemaker they neglected evanescent waves which are known to contribute to an important part of the free wave emission \citep{hudspeth-stokes-91}. \citet{tomita_89} found also the sine solution for $|B_4|$ contained in equation \eqref{eq:B4-mag-sol} without linking it however to the detuning sinc behavior first observed experimentally by \citet{LHS66} and \citet{McGoldrick66}.
\subsection{Waves in basins}
In the case of mechanically generated waves, the experiments show that the growing daughter wave has a frequency in exact resonance condition $\omega_4=2\omega_1-\omega_3$. The direction $\theta_4$ of the daughter-wave wavenumber ${\mathbf k_4}$ is still unknown ; the condition for wavenumbers may be not fullfilled and a mismatch or detuning can exist $\Delta {\mathbf k}=2{\mathbf k_1}-{\mathbf k_3}-{\mathbf k_4}$. Although the direction of the daughter wave is not specified, we assume that the fastest growing daughter wave is the one with minimal detuning. In other words, the daughter wave propagates along the direction of $2{\mathbf k_1}-{\mathbf k_3}$ and the corresponding detuning is now $\Delta k=\left|2{\mathbf k_1}-{\mathbf k_3}\right|-k(2\omega_1-\omega_3)$.

%With the Hamiltonian formalism, the relation between the space amplification factor $G$ at resonance used by \cite{LH62} and the interaction kernel $T_{1134}$ reads
Using equation \eqref{eq:B2eta} to convert from generalized to wave amplitudes, we obtain the relation between the interaction kernel $T_{1134}$ and the space amplification factor $G$ in \citet{LH62} 
$$
G  = \frac{T_{1134}}{k_1^3} \left(\frac{k_4}{k_1} \right)^{3/4}  \left(\frac{k_3}{k_1}\right)^{-5/4} {\rm \ .}
$$
%$$
%G = \frac{\omega_4}{\omega_1} \, \frac{T_{1134}}{k_1^3} \left(\frac{k_4}{k_1}\right)^{1/4} \left(\frac{k_3}{k_1}\right)^{-5/4}{\rm \ .}
%$$
%We have noticed a misprint in equation (6.4) in \citet{LH62}: the term $-(6+\xi^2)^{1/2}$ should be replaced by $-{\rm sgn}(\xi)(6+\xi^2)^{1/2}$ where $\xi=(1-r)/r$.
%
\subsection{Large amplitude}
At or near resonance and after a long enough time the daughter wave may reach large amplitude and our assumption $|B_4|\ll|B_1|$ and $|B_3|$ becomes invalid. In that case the exact analytical solution is expressed by means of Jacobian elliptic functions (see \citet{Stiassnie2005} for instance). The idea or concept of exact resonance must be limited to only the initial stage. Furthermore, no exact nonlinear resonance conditions can exist since the total phase evolves nonlinearly in time; the detuning $\Delta\Omega$ cannot stay null if a resonant transfer occurs as the amplitudes will evolve and then modify the detuning. In other words, all the four-wave interactions are off-resonance ones. The concept of exact resonance is meaningfull only at the initial stage. It corresponds to a linear growth of the daughter wave with maximum growth rate.

\bibliographystyle{jfm}

\bibliography{LH}